\documentclass[prb,
preprint,
amsmath,
amssymb,
showpacs,
superscriptaddress]{revtex4-1}
\usepackage{graphicx} 
\usepackage{dcolumn}  
\usepackage{bm}       
\usepackage{amsfonts}
\usepackage{dsfont}
\usepackage{csquotes}

\usepackage{ulem}
\usepackage{color}

\usepackage{float}
\usepackage{physics}
\usepackage{comment}
\renewcommand{\k}{\mathbf{k}}
\begin{document}

\title{Role of topological charges in the nonlinear-optical response from Weyl semimetals}

\author{Amar Bharti}
\affiliation{%
Department of Physics, Indian Institute of Technology Bombay,
           Powai, Mumbai 400076, India }

\author{Gopal Dixit}
\email[]{gdixit@phy.iitb.ac.in}
\affiliation{%
Department of Physics, Indian Institute of Technology Bombay, Powai, Mumbai 400076, India }
\affiliation{%
Center for Computational Sciences, University of Tsukuba, Tsukuba 305-8577, Japan}

\date{\today}


\begin{abstract}
The successful realization of the topological Weyl semimetals has revolutionized contemporary physics. 
In recent years, multi-Weyl semimetals, a class of  topological Weyl semimetals,  has 
attracted broad interest in condensed-matter physics. 
Multi-Weyl semimetals are emerging topological semimetals with nonlinear anisotropic energy 
dispersion, which is characterized by higher topological charges. 
In this study, we investigate how the topological charge affects the nonlinear optical response 
from multi-Weyl semimetals. 
It has been observed that the laser-driven electronic current is characteristic of the topological charge, and
 the laser polarization's direction influences the current's direction and amplitude. 
In addition, the anomalous current, perpendicular to the laser's polarization, 
carries a distinct signature of the topological charges  
and encodes the information about the parity and amplitude of the nontrivial Berry curvature.
We show that the anomalous current associated with the anomalous Hall effect 
remains no longer proportional to the topological charge  
at higher laser intensity --  a significant deviation from the linear response theory. 
High-harmonic spectroscopy is employed to capture the distinct and interesting features of the currents in multi-Weyl semimetals where the topological charge drastically impacts the harmonics' yield and energy cutoff. 
\end{abstract}

\maketitle

\newpage

\section{Introduction}
Recent years have witnessed a proliferation in studies of  quantum materials, such as 
topological insulators, Dirac and Weyl semimetals, as these materials 
are not only crucial for fundamental research 
but also have immense applications in upcoming quantum technologies.  
Owing to the presence of the topologically protected Weyl points,  
Weyl semimetals (WSMs) have become a focal point among the quantum materials~\cite{armitage2018weyl,yan2017topological,vazifeh2013electromagnetic}. 
Isolated points  in momentum space 
at which valance and conduction bands, with linear energy dispersions, 
touch  are known as  Weyl points. 
These points  appear in pairs with  opposite chirality  and can be visualized 
as the source and sink of the Berry curvature.
The low-energy collective excitations in  WSMs are  massless Weyl fermions,
which make an elegant connection between high-energy and condensed-matter physics.

The three-dimensional Dirac semimetals possess 
both time-reversal and inversion protection, giving four fold degenerate Dirac points.
If one of these two symmetries is broken, then these Dirac points transform into Weyl points. 
The absence of the time-reversal or inversion symmetry in the WSMs  gives rise to  
the Berry curvature.
The topology of the Berry curvature in WSMs results  in a series of 
exotic phenomenons, such as the chiral magnetic effect ~\cite{vazifeh2013electromagnetic, li2016chiral,kaushik2019chiral}, the circular photogalvanic effect ~\cite{de2017quantized,ma2017direct,rees2020helicity,le2021topology}, and the anomalous Hall effect ~\cite{shekhar2018anomalous,meng2019large, yang2015chirality} to name but a few~\cite{burkov2014chiral, trescher2015quantum, kim2013dirac}.
Interaction of light with WSMs
has been a frontier approach to probe various exotic phenomena associated with the Berry curvature in 
WSMs~\cite{lv2021experimental, lv2021high, bao2021light, orenstein2021topology, dantas2021nonperturbative, tamashevich2022nonlinear, nathan2022topological, matsyshyn2021rabi,  ahn2017optical, sirica2021shaking}.
The WSMs also hold promise for future technologies based on the ultrafast photodetector, the 
chiral terahertz laser source, and quantum physics~\cite{osterhoudt2019colossal,gao2020chiral}.

Earlier realizations of  the WSMs have a  unit topological charge,  $n =1$ ~\cite{lv2015experimental,xu2015discovery,morali2019fermi,liu2019magnetic,belopolski2019discovery}. 
The sign of the  Berry-curvature  monopole determines the sign of the topological charge of the  WSMs. 
The magnitude of the topological charge is $n = \int_{BZ} \nabla_{\mathbf{k}}\cdot\Omega(\mathbf{k})~ d\mathbf{k}$, with 
 $\Omega(\mathbf{k})$ being  the Berry curvature. 
A class of  WSMs, multi-Weyl semimetals (m-WSMs), with  
topological charges higher than one came into existence in recent years.  
The m-WSMs with topological charges two and three are, respectively,  manifested by 
quadratic and cubic energy dispersions along particular directions, whereas they exhibit linear dispersion along other directions~\cite {fang2012multi,huang2016new, xu2011chern,liu2017predicted}. 
This contrasts with ``conventional'' WSMs with   topological unit charges, 
which exhibit isotropic linear energy dispersion. 
Weyl points in m-WSMs can be formed by 
annihilating Weyl points of the same chirality in WSMs with lower topological charges~\cite{roy2022non}. 
The upper bound on the topological charge in m-WSMs is limited to three by 
certain crystalline symmetries~\cite{yang2014classification, fang2012multi, xu2011chern}. 
A few possible potential candidates for the m-WSMs are 
SiSr$_2$ and HgCr$_2$Se$_4$~\cite{fang2012multi,huang2016new, xu2011chern}.

The m-WSMs  manifest unique quantum response due to the higher topological charges and resultant anisotropic energy dispersions~\cite{nag2020thermoelectric}. 
It has been shown that the chirality accumulation is possible without a magnetic field 
due to higher topological charges~\cite{huang2017topological}.
A distinct chiral anomaly-induced nonlinear Hall effect, associated with the 
different topological charges in m-WSMs, has been discussed~\cite{nandy2021chiral}. 
Moreover, transport properties are significantly altered in comparison to the single-WSMs~\cite{menon2020anomalous, menon2021chiral, nag2022distinct, chen2016thermoelectric, gorbar2017anomalous, park2017semiclassical}. 
It has been found that the anomalous Hall current  in the nonperturbative regime 
saturates, and the anomalous Hall conductivity scales linearly with the topological charges~{\cite{dantas2021nonperturbative,nandy2019generalized}.  
By employing the effective field theory, 
the signature of a non-Abelian anomaly in m-WSMs has been investigated~\cite{dantas2020non}.

In this work, we investigate how the nonlinear optical responses in m-WSMs  
are sensitive to the topological charges for different laser intensities.   
In the following, we show that  the  current, parallel to the laser's polarization, and 
the anomalous current, perpendicular to the laser's polarization, exhibit distinct behaviour as a function 
of the laser's intensity for different topological charges. 
In addition, we demonstrate that the  current is anisotropic in nature, which strongly depends 
on the direction of the laser polarization. 
Moreover,  the anomalous current is sensitive to  the line connecting the Weyl points. 
The  parallel current increases linearly below a critical laser's intensity. 
However, above a critical intensity, the parallel current 
shows a nonlinear increment for different topological charges, and it starts saturating after another critical intensity. 
On the other hand, the anomalous current displays linear behavior for relatively larger intensity.  
The Berry-curvature driven anomalous current approaches  similar values for 
different topological charges at large intensity limits. 
We employ high-harmonic spectroscopy to probe the distinct behaviors of the normal and anomalous currents in m-WSMs. 
It has been found that the topological charge drastically alters the harmonics' yield and energy cutoff.

\section{Theoretical Framework}

Interaction of a m-WSM with an ultrashort intense laser pulse is simulated by solving 
density-matrix-based semiconductor Bloch equations in the Houston basis~\cite{mrudul2021light} as 
\begin{subequations}\label{eq:sbe}
\begin{align}
 \pdv{\rho_{cv}^\mathbf{k}}{t} & = \left[ -i\epsilon_{cv}^{\mathbf{k}_t} -\frac{1}{\textrm{T}_{2}}\right] 
 \rho_{cv}^{\mathbf{k}} + i\mathbf{E}(t)\cdot  \mathcal{D}_{cv}^{\mathbf{k}_t}  \left[\rho_{vv}^{\mathbf{k}}-\rho_{cc}^{\mathbf{k}}\right] \\
 \pdv{\rho_{vv}^\mathbf{k}}{t}  & = i\textbf{E}(t)\cdot \mathcal{D}_{vc}^{\textbf{k}_t}~\rho_{cv}^{\textbf{k}} + \textrm{c.c.}
\end{align}
\end{subequations} 
Here, $\epsilon_{cv}^{\mathbf{k}_t}$ and $\mathcal{D}_{cv}^{\mathbf{k}_t}$ are, respectively, 
energy band gap  and dipole matrix elements between  conduction and valence  bands at $\mathbf{k}_{t}$.
In the presence of an intense laser, the crystal momentum changes from $\mathbf{k}$ to $\mathbf{k}_{t}$ as 
$\mathbf{k}_t=\mathbf{k} + \mathbf{A}(t)$, where $\mathbf{A}(t)$ as  the vector potential of the laser and  
is related to the laser's electric field $\mathbf{E}(t)$  as $\mathbf{A}(t) = - \int_{-\infty}^{t} \mathbf{E}(t^{\prime}) dt^{\prime}$. 
A phenomenological term $\textrm{T}_{2}$ is introduced to account for the decoherence between the hole and the electron during  the light-matter interaction process. 
Fourth-order Runge-Kutta method is used to solve Eq. \eqref{eq:sbe} numerically for $\rho$ at each 
time step. 
The completely filled valence band ($\rho_{vv} = 1$) and the empty conduction band ($\rho_{cc} = 0$) are used as initial conditions. 
The dipole matrix element and the energy are obtained from eigenstates and eigen energies by the diagonalizing Hamiltonian given in Eqs.~(\ref{eq:n_1}-\ref{eq:n_3}) as discussed in Ref.~\cite{mrudul2021high}. }

The total current is evaluated as 
$\mathbf{J}(\mathbf{k},t) =  \sum_{m,n  \in \{c,v\} } \rho_{mn}^{\mathbf{k}} \mathbf{p}_{nm}^{\mathbf{k}_t}$
with $\mathbf{p}_{nm}^{\mathbf{k}_t}$ being the momentum matrix-element. 
The total current is decomposed into two components: intraband and non-intraband currents. 
The intraband current is solely originating due to group velocity and can be expressed  as  
$\mathbf{J}_{intra}(\mathbf{k},t)=\sum_{n}\nabla_{\mathbf{k}}\epsilon_n^{\mathbf{k}_t}\rho_{nn}^\mathbf{k} $. On the other hand, the non-intraband current contains the contribution due to the Berry curvature, among others~\cite{yue2022introduction}. 
The Berry curvature is expressed in terms of  the curl of the Berry connection: $\mathbf{\Omega}= \nabla \times \mathcal{A}$.  
The Berry connection  between conduction and valence bands  
can be evaluated  as
$\mathcal{A}_{cv}(\mathbf{k}) = -i \braket{\nabla_{\mathbf{k}} u_{c}({\mathbf{k}})}{u_{v}({\mathbf{k}})} = \mathcal{A}_{vc}^{*}(\mathbf{k})$, which is visualized as  the off-diagonal dipole matrix elements~\cite{nematollahi2020topological,ngo2021microscopic}.
Thus, the  Berry curvature between conduction and valence bands is written as 
$\mathbf{\Omega}_{cv}(\mathbf{k}) = i (\mathcal{A}_{vc} (\mathbf{k}) \times \mathcal{A}_{cv}(\mathbf{k}))$. 
The current due to the Berry curvature is evaluated as  $\mathbf{J}_\mathbf{\Omega}(\mathbf{k},t) = \mathbf{E}(t) \times \sum_{cc,vv} \mathbf{\Omega}_{cv}(\mathbf{k})~\rho_{cc,vv}(\mathbf{k}, t)$ 	
and is commonly known as the anomalous current.

Inversion-symmetric m-WSM with broken time-reversal symmetry for different 
topological charges $n$ can be collectively written as 
$\mathcal{H}^{(n)}(\mathbf{k}) = \mathbf{d}^{(n)}(\mathbf{k}) \cdot \sigma$, 
with $\sigma$'s being the Pauli matrices and $\mathbf{d} = [d_{x}, d_{y} ,d_{z}]$~\cite{dantas2020non}.
The full expressions of  the three components of $\mathbf{d}^{(n)}(\mathbf{k})$ for  topological charges $n = 1, 2,$ and 3 are  
written as  	 
\begin{eqnarray}\label{eq:n_1}
 \mathbf{d}^{(1)}(\mathbf{k}) & =  & \left[t\sin(k_x a), t\sin(k_y a)\right.,\\\nonumber
&&\left. t\{\cos(k_z a) - \cos(k_0 a) +2- \cos(k_x a) - \cos(k_y a)\}\right],
\end{eqnarray}
\begin{eqnarray}\label{eq:n_2}
\mathbf{d}^{(2)}(\mathbf{k}) & = & \left[t\{\cos(k_x a)-\cos(k_y a)\}, t\sin(k_x a)\sin(k_y a),\right. \\\nonumber
&&  \left. t\{\cos(k_z a) - \cos(k_0 a) +2- \cos(k_x a) - \cos(k_y a)\}\right],
\end{eqnarray}
and 
\begin{eqnarray}\label{eq:n_3}
\mathbf{d}^{(3)}(\mathbf{k}) & = & \left[t\sin(k_x a)\{3\cos(k_y a)-\cos(k_x a) -2\},\right.\\\nonumber
&&\left. t\sin(k_y a)\{3\cos(k_x a)-\cos(k_y a) - 2\}, \right.\\\nonumber
&&\left. t\{\cos(k_z a) - \cos(k_0 a) +2- \cos(k_x a) - \cos(k_y a)\}\right].
\end{eqnarray}
Here, $k_0$ determines the positions of the Weyl nodes, which 
twe have taken to be $k_0 = \pi/(2a)$ in this article throughout, unless stated otherwise. 
Thus, the positions of Weyl points for Eqs.~(\ref{eq:n_1}-\ref{eq:n_3}) are $(0,0,\pm k_0)$ as shown in 
Fig.~\ref{fig:Fig1}.

\section{Results and Discussion}

\begin{figure}[!hbt]
\includegraphics[width=0.7\linewidth]{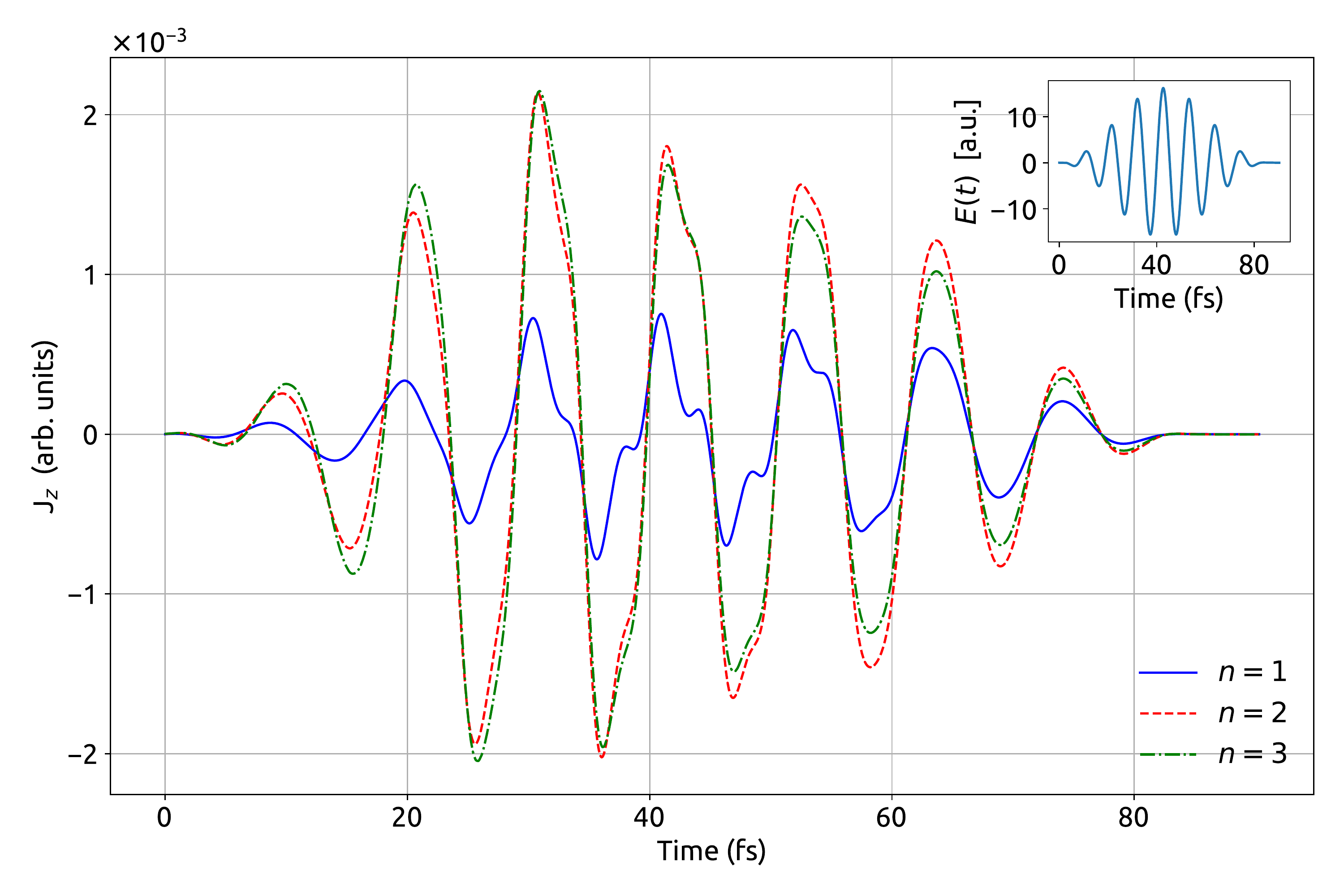}
\caption{Total current along the $z$ direction in multi-Weyl semimetals for different 
topological charges $n$. 
The linearly polarized laser pulse along the $z$ direction is  employed to generate the current and  is 
shown in the inset. 
The pulse is approximately 100 fs long with a  wavelength of $3.2~\mu$m and 
an intensity of  $10^{11}$ W/cm$^2$. 
The value of the phenomenologically  decoherence time $\textrm{T}_{2}$ is 1.5 fs.}\label{fig:Fig2}
\end{figure}

Let us apply  a laser polarized along the $z$ direction to induce electron dynamics in m-WSMs. 
Figure~\ref{fig:Fig2} presents the total current in m-WSMs with different  topological charges. 
The current in a WSM with $n = 2$ is approximately one-order higher in magnitude  
in comparison to the current for $n = 1$. 
However, the total current  is comparable for $n = 2$ and 3 as evident from  Fig.~\ref{fig:Fig2}.
The overall shape of the currents in all cases  are significantly distorted with respect to the shape
of the laser's electric field as shown in the inset. 
This distortion indicates  the involvement of nonlinear optical processes during  electron dynamics.  
Note that  the laser polarized along the $z$ direction does not yield current along any direction other than the polarization direction, which will be discussed later.

To understand the unusual behavior in the strength of the total current for different $n$ values, let us first analyze
the energy  band structures of the m-WSMs.     
Owing to a similar structure of $\mathcal{H}^{(n)}(\mathbf{k})$, the eigenvalues can be succinctly written as $E_{c,v}= \pm \sqrt{d_{x}^{2} + d_{y}^{2} + d_{z}^{2}}$, where $E_{c,v}$ corresponds to conduction and valence bands, respectively;  and the $d$'s are corresponding $\mathbf{d}$'s for each $n$ as given in 
Eqs.~(\ref{eq:n_1} - \ref{eq:n_3}).

\begin{figure}[!hbt]
\includegraphics[width=0.7\linewidth]{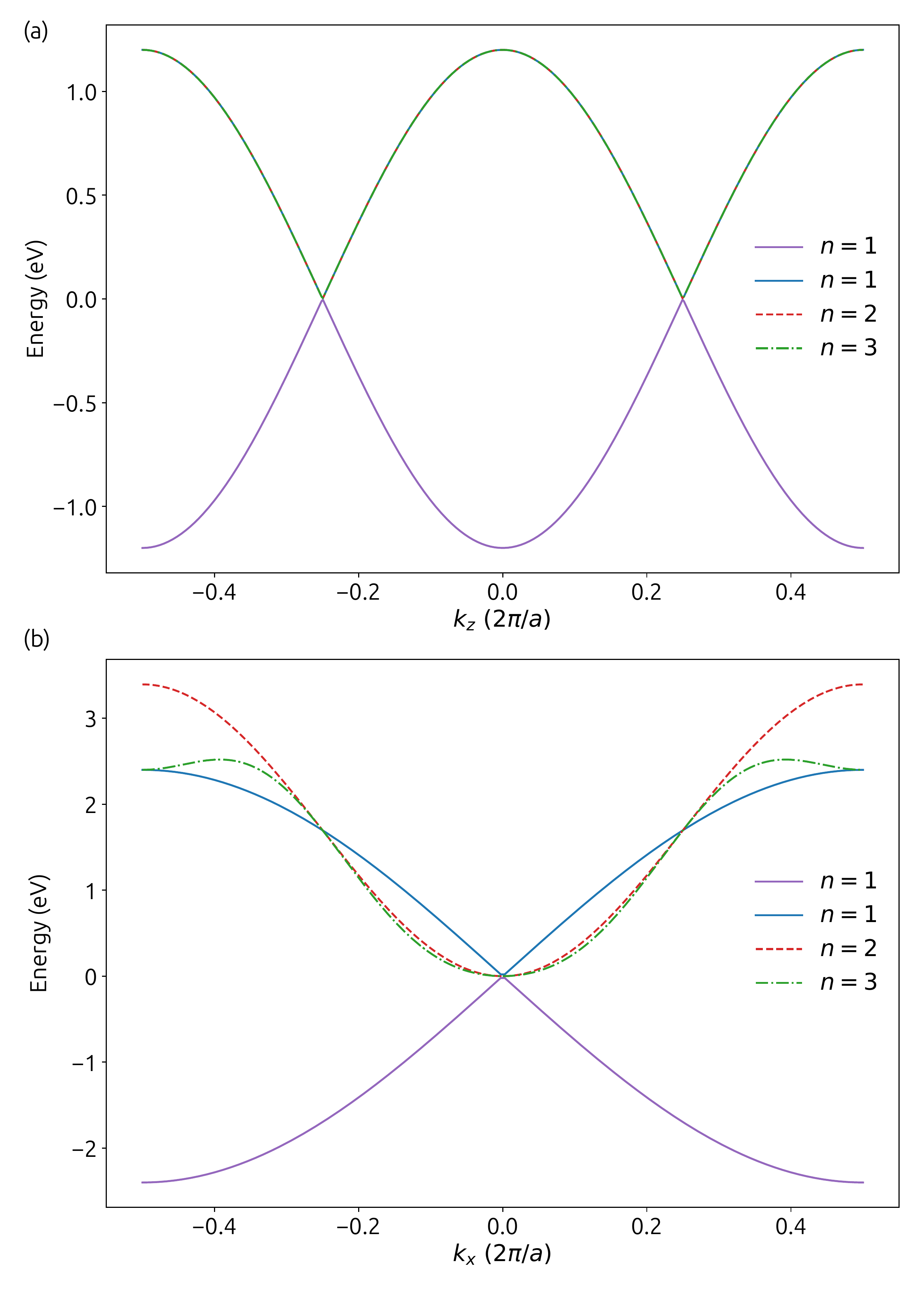}
\caption{Energy band structure corresponding to multi-Weyl semimetals with different topological 
charges along (a) the $k_{z}$ direction, and (b) the $k_{x}$  direction around a Weyl node. 
The band structures along the $k_{y}$ and  $k_{x}$ directions are identical.
The band structure is obtained by diagonalizing the Hamiltonian $\mathcal{H}^{(n)}$   for different 
values of $n$ with the hopping parameter  $t = 1.2$ eV and the lattice parameter $a = 6.28$ \AA. 
The purple line corresponds to the valence band in all cases.}
\label{fig:Fig1}
\end{figure}

Figure~\ref{fig:Fig1}(a) shows the energy band structures along the $k_{z}$ direction for different $n$ values.  
The band structures  with two Weyl nodes of opposite chirality at  $\pm k_0$ 
are identical for $n = 1, 2$, and 3 as evident from the figure. 
The reason behind the identical energy dispersion along $k_{z}$  can be attributed to the 
same hopping parameter $t$ and lattice parameter  $a$ for different $n$'s. 
However, the band structures  near one of the Weyl nodes 
are significantly different  along $k_{x}$ for  different $n$'s as documented in Fig.~\ref{fig:Fig1}(b). 
Note that the band structures  along  $k_{x}$ and $k_{y}$ are identical. 
The different energy  dispersions along $k_{x}$ are directly related to the topological charges 
of the Weyl nodes. 
The relation between the energy dispersion and the topological charge is apparent from  
the low-energy Hamiltonian near a Weyl node as 
$\mathrm{H}^{(n)}(\k) = v\left(k_x^n \sigma_x + k_y^n\sigma_y + k_z \sigma_z\right)$ for $n=1, 2$, 
and 3. 
It is straightforward to see that  
the effective low-energy band structures are linear, quadratic, and cubic for $n =1$, 2, and 3, 
respectively. 
Furthermore, it can be shown explicitly by calculating Berry curvature 
that the Chern numbers corresponding to linear, quadratic and cubic  dispersions are one, two, and three, respectively, as calculated elsewhere~\cite{nag2022distinct}.

Analysis of  Fig.~\ref{fig:Fig1}  indicates that the current should be comparable for 
the m-WSMs with different $n$'s as the energy dispersions are identical along the direction of laser polarization.  
However, the energy gap between valance and conduction bands is smaller for $n = 2$ and 3 in comparison 
to $n =1$ around the Weyl node as evident from the energy dispersion along $k_{x}$, which translates 
into a higher probability of electron excitation for $n = 2$ and 3. 
Moreover, the energy dispersions for $n = 2$ and 3 are similar.
The distinct energy dispersion naturally leads to the distinct gradient of the group velocity. 
Thus, the cubic band dispersion offers a larger intraband current  compared to the quadratic one. 
Similarly, the linear dispersion has the smallest current flow out of the three. 
Both these facts explain why the total current exhibits similar features 
for $n = 2$ and 3 and is higher in magnitude in comparison to $n =1$.   
Note that the entire Brillouin zone contributes to the total current in three-dimensional systems like WSMs~\cite{gu2022full}.  
  
\begin{figure}[!hbt]
\includegraphics[width=0.7\linewidth]{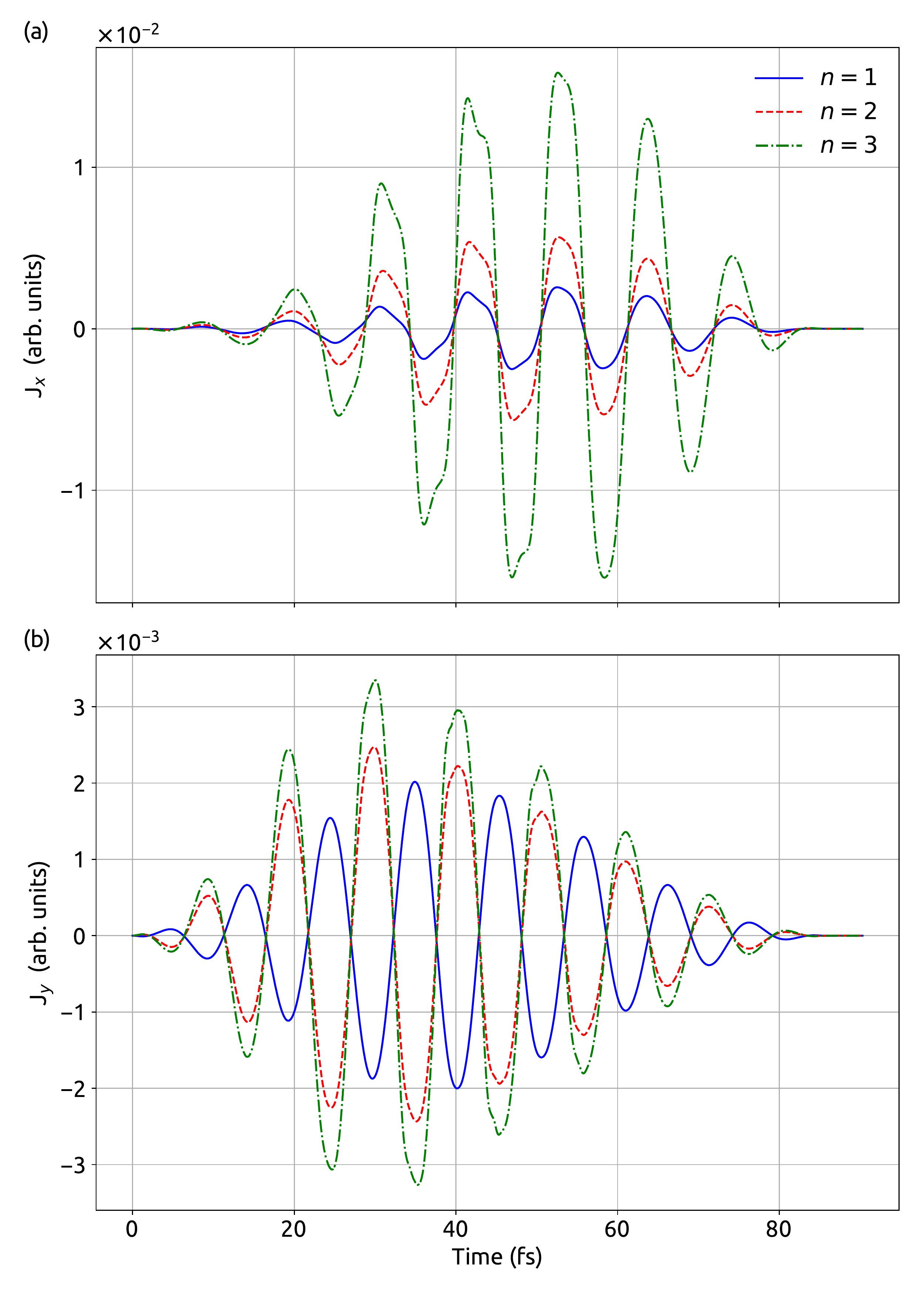}
\caption{Total current along (a) the direction of the laser polarization, i.e., normal current, and (b) 
along the perpendicular $y$ direction,  i.e., anomalous current.  
The laser is polarized along the $x$ direction. 
Details of the simulation  and the laser parameters are the same as those in Fig.~\ref{fig:Fig2}.}
\label{fig:Fig3}
\end{figure}

So far, we have witnessed that the laser polarized along the $z$ direction does not generate current along any 
perpendicular direction. 
Let us see how this observation is modified  when the polarization of the laser is tuned  
from the $z$ direction to the $x$ direction. 
Figure~\ref{fig:Fig3}(a) presents the current parallel to the laser polarization for different 
topological charges.  The strength of the current is drastically different for different $n$'s, which
is expected from the band structures along $k_{x}$ as shown in Fig.~\ref{fig:Fig2}(b). 
The current increases monotonically as the band dispersion goes from linear to quadratic and cubic with 
the increase in  $n$. 
This observation is in contrast with the previous one where the current was comparable for $n = 2$ and 3.   
Before we delve into the detailed reason for the increase in current as  a function of $n$, 
let us focus on the perpendicular component of the total current along the $y$ direction 
as shown in Fig.~\ref{fig:Fig3}(b). 
A similar perpendicular component appears along the $x$ direction when the laser is polarized along the $y$ direction. 
In the following, let us use the terminology normal and anomalous currents for the currents along and perpendicular to the laser polarization, respectively, as used in a previous study~\cite{bharti2022high}.

We employ a semiclassical equation of motion for laser-driven electron dynamics 
to understand why the $x$ polarized laser generates  anomalous current, whereas $z$ polarization does not. 
It is known that the WSM has a non zero Berry curvature, which results  
an anomalous velocity as $\mathbf{E}(t) \times \mathbf{\Omega}$. 
Moreover, the Berry curvature in an  inversion-symmetric WSM follows 
$\mathbf{\Omega}(-\k) = \mathbf{\Omega}(\k)$, which leads  to nonzero current due to the anomalous velocity of  $\int_{\k} \rho(\k) \{\mathbf{E}\times \mathbf{\Omega}(\k)\}~d\k$. 
The parity of the Berry curvature's components in the present case [see Supplemental Material  i for details~\cite{NoteX}] is 
such that there is no anomalous velocity  if the electric field is along the 
same direction as the line connecting the Weyl nodes as discussed in Ref.~\cite{bharti2022high}.
Moreover, the anomalous current in the time-reversal broken WSM produces  the 
anomalous Hall effect~\cite{xiao2010berry}. 
Thus, the anomalous current is zero when the laser is polarized along  the $z$ direction, 
the direction along which the two Weyl nodes are situated. 
The situation changes drastically as the polarization direction changes from $z$ to $x$ or $y$.  

The anomalous current is proportional  to $\mathbf{\Omega}$, which means it is also proportional to $n$. 
Thus, the strength of the anomalous current increases as $n$ increases [see Fig.~\ref{fig:Fig3}(b)]. 
However, the anomalous currents due to $n = 2$ and 3 are out of phase with respect to $n = 1$. 
This behavior is due to  the sign of the integral of the Berry curvature's components as shown in Ref.~\cite{bharti2022high}. 
The sign of $\int_{\k} \{\mathbf{E}\times \mathbf{\Omega}(\k)\}~d\k$
is positive for $n$ = 1 and 3 and negative for $n$ = 2, which leads the anomalous
 current for $n$ = 2 out of phase.
Thus, the strength and the phase of the  anomalous current encode the information about the nontrivial topology of  the Berry curvatures in m-WSMs. 
Note that the anomalous current is one-order weaker  than the normal current as evident from  Fig.~\ref{fig:Fig3}. 
At this juncture, it is interesting to wonder how these features in the normal and anomalous currents 
alter with the laser's intensity.  

\begin{figure}[!hbt]
\includegraphics[width=\linewidth]{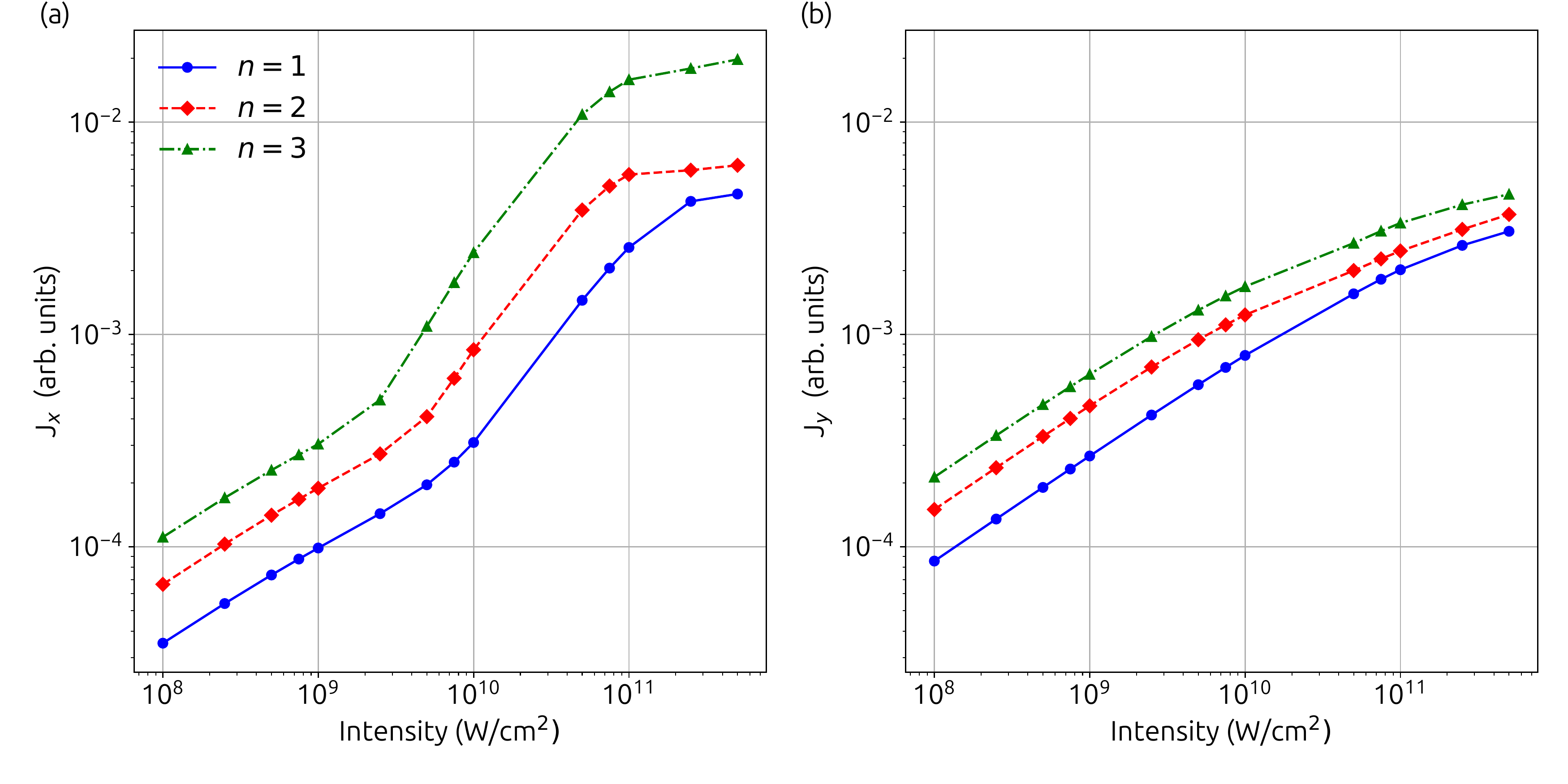}
\caption{Variations of the maximum amplitude for (a) the normal current along $x$ and (b) the anomalous current along $y$ as a function of the laser's intensity for multi-Weyl semimetals with topological charges $n = 1, 2,$ and 3. The other parameters are the same as those in Fig.~\ref{fig:Fig3}.}
\label{fig:Fig4}
\end{figure}

\begin{figure}[!hbt]
\includegraphics[width=0.7\linewidth]{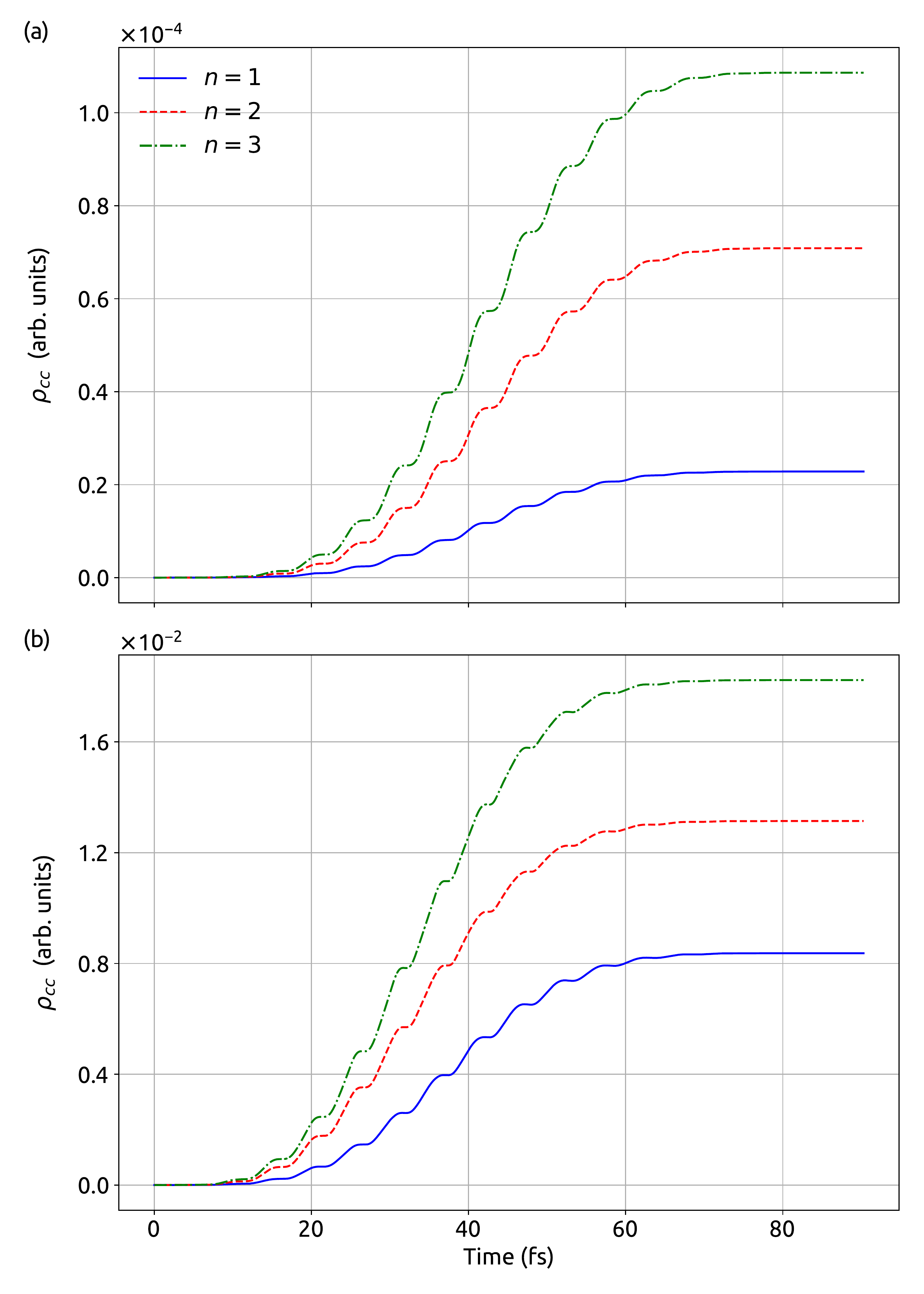}
\caption{Temporal evolution of the conduction band population, $\rho_{cc}$,  for different $n$  values 
during the laser for two intensities:  (a) $10^8$ W/cm$^2$ and (b) $10^{11}$ W/cm$^2$. The other parameters are the same as those in  Fig.~\ref{fig:Fig3}.}
\label{fig:Fig5}
\end{figure}

Figure~\ref{fig:Fig4} presents variation in the peak current for different $n$ values as a function of the intensity ranging from $10^8$  to $10^{11}$ W/cm$^2$. 
In the beginning, the anomalous current dominates over the normal current for each $n$ at  $10^8$ W/cm$^2$. 
However,  the normal current takes over the anomalous current  at some critical intensity.  
On comparing the normal and anomalous currents for each $n$, we find that this  critical intensity 
gets lower as  $n$ increases.  
In addition, the normal current starts to grow exponentially  
at a much lower intensity for higher $n$ values. 
The  peak current increases linearly as the intensity increases from $10^8$ to $10^{10}$ W/cm$^2$ [see Fig.~\ref{fig:Fig4}(a)].   
Moreover, the rate of increase is much higher for the normal current as compared to the anomalous current, which 
starts to saturate at an intensity much lower than that of the 
 normal current. 
There is no exponential growth region in the anomalous current compared to the normal current, which grows exponentially in the intensity window of $10^{10}$ to $10^{11}$ W/cm$^2$ [see Fig.~\ref{fig:Fig4}(b)].
It is expected that the laser drives electrons further away in the energy band as the intensity increases, 
and therefore the normal current increases. 
However, the comparatively large anomalous current at lower intensity is interesting.
Moreover, the subdued increment in the anomalous current needs further investigation. 
To understand  these interesting observations, we analyze the laser-driven electronic population in the conduction band. 

\begin{figure}[!hbt]
\includegraphics[width=0.7\linewidth]{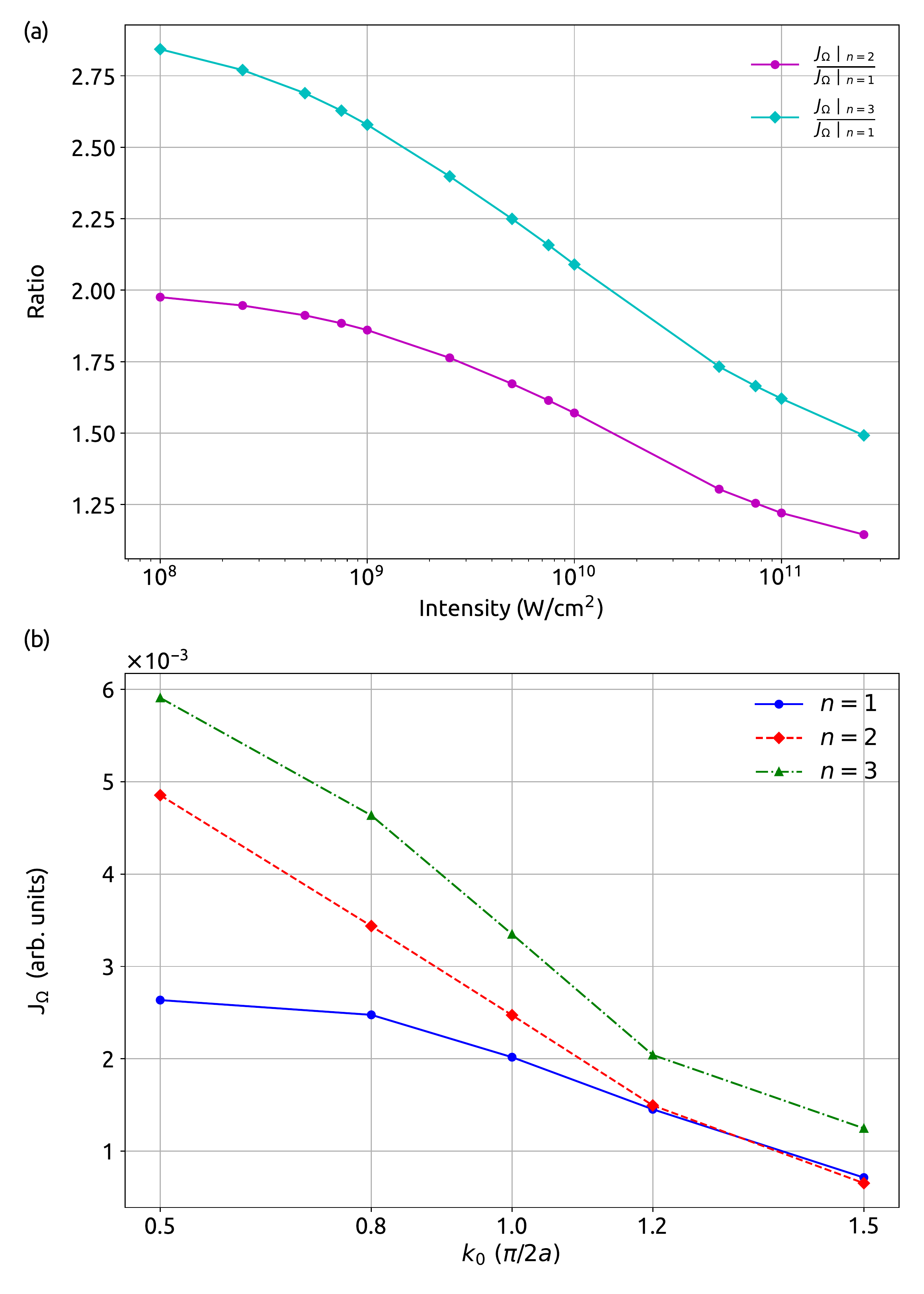}
\caption{(a) Ratio of the anomalous current's peak, $\textrm{J}_{\Omega}$, 
as a function of the laser's intensity, and 
(b) variation in the anomalous current's peak, $\textrm{J}_{\Omega}$,  
as a function of the separation between two Weyl nodes, $k_{0}$, at 
intensity $10^{11}$ W/cm$^2$ for different topological charges $n$.}
\label{fig:Fig6}
\end{figure}

Electronic population in the conduction band for  $n = 1$  is one oder in magnitude weaker  than 
the population  for  $n = 3$ as shown in Fig.~\ref{fig:Fig5}(a). 
Moreover, the conduction band  is sparsely populated for  $n = 1$ at $10^8$ W/cm$^2$.
As the intensity increases from $10^8$ to $10^{11}$ W/cm$^2$, the overall  
population increases by two orders in magnitude [see Fig.~\ref{fig:Fig5}]. 
The conduction band population for  $n = 3$ remains substantially more significant than that of the 
lower topological charges 
even at $10^{11}$ W/cm$^2$ as evident from Fig.~\ref{fig:Fig5}(b).
Similarly, the overall nature of the  populations remains 
the same and the difference in populations  is also pronounced for  different $n$ values 
at $10^9$ W/cm$^2$ (not shown here). 
There are  two key factors governing   the overall behavior of the  conduction band population for different $n$ values at two intensities: 
the first one is the reduction in the energy gap between valence and conduction bands around a Weyl node
as $n$ increases [see Fig.~\ref{fig:Fig1}(b)]. This results in higher probability of excitation and thus more population as $n$ increases.     
The other important factor is the dipole matrix amplitude and its relation with $n$ [see Eq. (\ref{eq:sbe})], 
which manifests higher electronic population as $n$ increases. 
Thus, it can be concluded from Fig.~\ref{fig:Fig5} that the population increases monotonically with 
$n$ irrespective of the laser's intensity.

Returning back to the considerable anomalous current at the lowest intensity $10^8$ W/cm$^2$, 
let us explore how the topological charge impacts  the anomalous current.
By following the  analysis of the linear response from the WSM, 
the anomalous current is proportional to $n$
as $\mathbf{J}_\mathbf{\Omega} \propto n (\mathbf{b}\times\mathbf{E})$, 
where $\mathbf{b}$ is
 the vector joining the Weyl nodes~\cite{nandy2019generalized}. 
The quantity $\mathbf{b}\times\mathbf{E}$ determines the direction of the anomalous current.
Note that the same reasoning
we have used earlier to explain why the laser polarized along the direction of the line connecting the 
Weyl nodes results  in no  anomalous current. 
If we consider the expression of $\mathbf{J}_\mathbf{\Omega}$ to be true for the considered laser intensities    
then,  the anomalous current should increase monotonically with intensity as 
$\mathbf{J}_\mathbf{\Omega} \propto \mathbf{E}$. 
However, the anomalous current  deviates significantly from this expectation  at higher intensity, signaling a nonlinear optical response from m-WSMs. 
It is important to emphasis 
that the anomalous current  along the perpendicular direction is mainly driven by the Berry curvature as shown in Fig. S1~\cite{NoteX}.

To corroborate our claim about the nonlinear optical response, 
let us analyze the ratio of the anomalous current's peak for different $n$ values 
as a function of the laser's intensity.
The ratios of the  peak current for $n=3$ to $n=1$, and $n=2$ to $n=1$ are presented in
Fig.~\ref{fig:Fig6}(a). 
Within the linear response framework, it is expected that the ratios 
$\mathbf{J}_\mathbf{\Omega}|_{n = 3} /\mathbf{J}_\mathbf{\Omega}|_{n = 1}$ and 
$\mathbf{J}_\mathbf{\Omega}|_{n = 2} /\mathbf{J}_\mathbf{\Omega}|_{n = 1}$ should be 3 and 2, 
respectively as 
$ \mathbf{J}_\mathbf{\Omega} \propto  n$. 
This expectation is true at   $10^{8}$ W/cm$^2$ where the ratios are  
close to 3 and 2 as evident from Fig.~\ref{fig:Fig6}(a). 
However, both ratios decrease monotonically as intensity increases, albeit at different rates. 
The ratios $\mathbf{J}_\mathbf{\Omega}|_{n = 3} /\mathbf{J}_\mathbf{\Omega}|_{n = 1}$ and 
$\mathbf{J}_\mathbf{\Omega}|_{n = 2} /\mathbf{J}_\mathbf{\Omega}|_{n = 1}$ reach approximately  2 and 1.5 at   
$10^{10}$ W/cm$^2$, respectively --  a drastic deviation from the expectation of linear response theory. 
The ratio $\mathbf{J}_\mathbf{\Omega}|_{n = 3} /\mathbf{J}_\mathbf{\Omega}|_{n = 1}$
decreases with a much faster rate compared to 
$\mathbf{J}_\mathbf{\Omega}|_{n = 2} /\mathbf{J}_\mathbf{\Omega}|_{n = 1}$. 
This implies that the anomalous  current due to the Berry curvature decreases faster with an increase 
in $n$. 
Thus, $\mathbf{J}_\mathbf{\Omega}$ originating from two different topological charges 
may be comparable at a certain intensity. 
However, it is practically not feasible to keep increasing 
the intensity as it can go above the damage threshold of the material. 

So far, we have investigated how the  anomalous current behaves with respect to the intensity.  
Recently, it has been shown that the nonlinear anomalous current  in a Weyl semimetal with $n =1$
decreases with increasing the distance between the Weyl nodes~\cite{bharti2022high,avetissian2022high}. 
Let us see how the reported observation changes for m-WSMs. 
Figure~\ref{fig:Fig6}(b) presents how the anomalous current's peak varies with  
the distance between the Weyl nodes for different $n$ values. 
The current due to the Berry curvature decreases with an increment in the distance between the nodes, 
and the decrease is different for each $n$. 
Similar observations can be made when the intensity lies within the linear-response regime as evident from Fig.~S2~\cite{NoteX}.
The overall behavior of the anomalous current  remains  same, but the response due to the change in the distance  
is nontrivial when we compare for different $n$ values.

\begin{figure}[!hbt]
\includegraphics[width=0.7\linewidth]{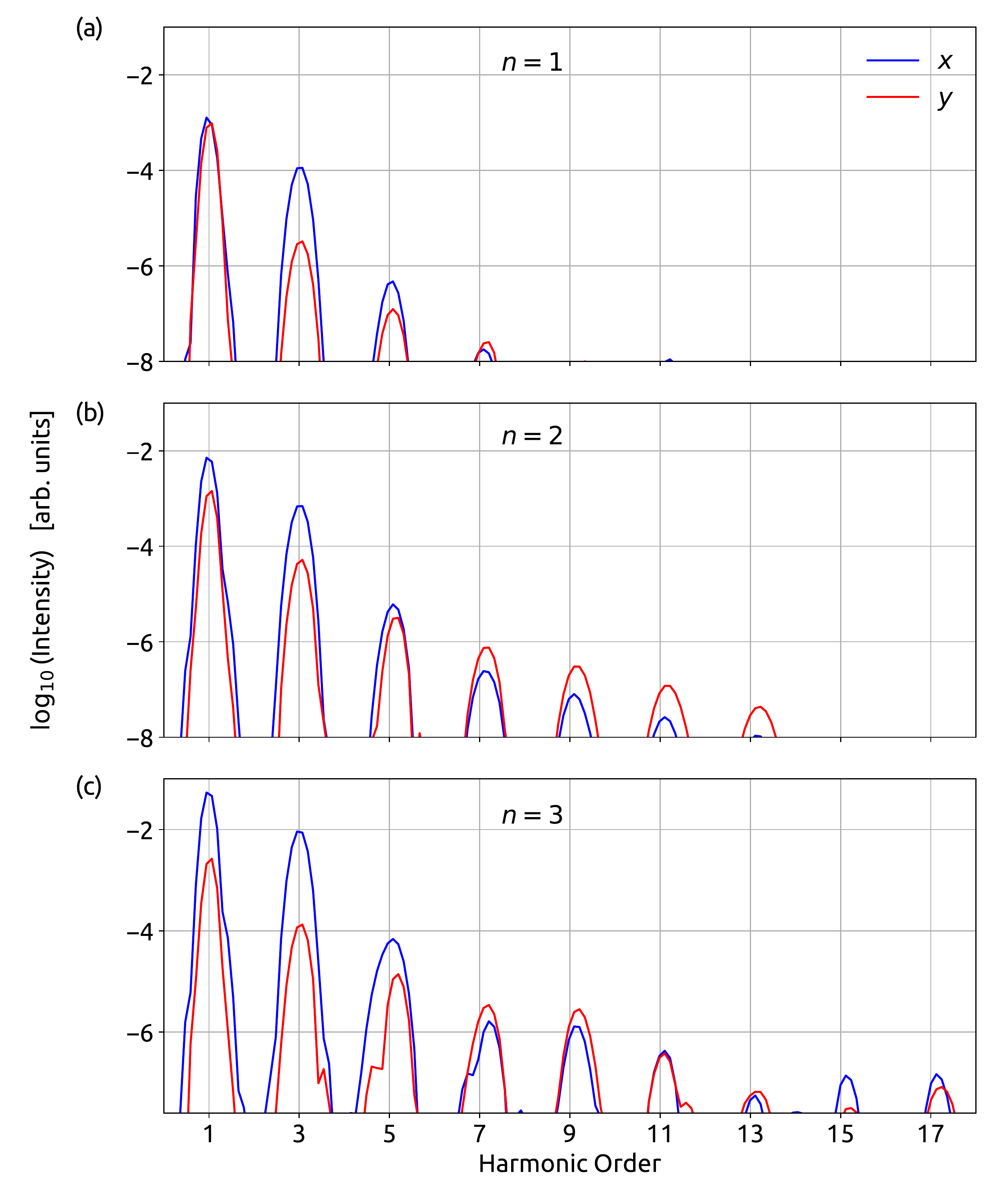}
\caption{Higher-order harmonic spectrum corresponding to multi-Weyl semimetals with topological charge 
(a) $n = 1$, (b) $n = 2$, and (c) $n =3 $.  The driving laser is 100 fs long 
with  wavelength of $3.2~\mu$m and 
intensity of  $1\times10^{11}$ W/cm$^2$. 
The laser pulse is linearly  polarised along $x$ direction. 
A phenomenologically  decoherence time $\textrm{T}_{2}$  of 1.5 fs is added in the simulation.
The harmonic spectrum is simulated by taking the modules square of the Fourier transform  of the time-derivative of the total current as $\left|\mathcal{FT}\left(\frac{d}{dt} \left[\int \textbf{J}(\mathbf{k}, t)~d{\textbf{k}} 
\right] \right) \right|^2$~\cite{mrudul2020high}. The distance between two Weyl node is  $k_0 = \pi/(2a)$ with $a = 6.28$ \AA.} \label{fig:Fig7}
\end{figure}

The normal and anomalous currents are analyzed as the optical responses in m-WSMs, which  
transit from a linear regime to a nonlinear regime as the laser's intensity increases, 
even from a perturbative regime to a nonperturbative regime.
High-harmonic generation (HHG) is a hallmark example of a nonperturbative nonlinear optical process, which has become a method of choice to probe various 
static and dynamical aspects of solids~\cite{schubert2014sub, mrudul2021light, mrudul2021controlling, hohenleutner2015real, zaks2012experimental, pattanayak2020influence,  pattanayak2019direct, pattanayak2022role, rana2022probing, rana2022high}.   
In addition,  HHG from topological materials has become a center  of attention as it allows one 
to investigate nonequilibrium topological aspects of  topological insulators, Dirac  
and Weyl semimetals in recent years~\cite{baykusheva2021all, kovalev2020non, cheng2020efficient, bai2021high, dantas2021nonperturbative, lv2021high}. 
Recently, it has been shown that  high-harmonic spectroscopy  can be used to probe the  
light-induced nonlinear anomalous Hall effect in a Weyl semimetal~\cite{bharti2022high}.
Thus, it is interesting to explore how the topological charge impacts HHG from m-WSMs. 

The higher-order harmonic spectra corresponding to m-WSMs are distinct for different $n$'s as can be seen  from Fig.~\ref{fig:Fig7}. 
Odd-order harmonics are only generated as m-WSMs exhibit 
 inversion symmetry in the present case. 
The harmonic cutoff not only increases drastically but 
also the yield of the  harmonics is boosted by several orders  as $n$ increases. 
The energy cutoff increases from seven to thirteen as $n$ changes from $n = 1$ to $n = 2$. 
Moreover, the yield of the seventh harmonic is two-orders in magnitude  boosted 
as $n$ transits from 1 to 2 as is evident from Figs.~\ref{fig:Fig7}(a) and \ref{fig:Fig7}(b).  
Similar observations can be made for the harmonics shown in Fig.~\ref{fig:Fig7}(c) for $n = 3$.
The Berry-curvature-driven anomalous current  in m-WSMs  
results in anomalous odd harmonics along the $y$ direction. 
The presence of the anomalous harmonics is a signature of the light-induced anomalous Hall effect, and 
the strength of these harmonics gives the measure of the Hall effect~\cite{bharti2022high}.

The characteristic dependence of the relative yield of the normal and anomalous harmonics on the harmonic-order 
offers a route to tailor the polarization of the emitted harmonics, which carry information about 
the topology of the Berry curvature in m-WSMs. 
The ellipticity of the first, third, and fifth harmonics for $n =1$ reads as  0.85, 0.17, and 0.55, respectively.
As the value of the topological charge changes to $n = 2(3)$, the ellipticity of the first, third, and fifth harmonics  changes drastically as 0.41(0.2), 0.25(0.11), and 0.72(0.45), respectively. 
The reduction of the ellipticity of a given harmonic with an increase in  the topological charge can be attributed to a significant change in the yield of the  anomalous harmonic. 
Thus,  the ellipticity of the emitted harmonic for a given  harmonic order is significantly different for different  $n$ values and  
can be potentially used as a characterization tool for the  topological charge in m-WSMs. 

\section{CONCLUSION}
In summary, we have explored the impact of the topological charge  
on laser-driven electron dynamics in  Weyl semimetals. 
For this purpose,  multi-Weyl semimetals with topological charges $n =1, 2$, and 3 are considered. 
It has been found that the laser-driven electronic currents are distinct for different $n$'s.   
Moreover, the direction and amplitude of the current strongly depend on the laser polarization. 
The  parity and amplitude of the  Berry curvature determine the behavior of the 
anomalous current -- current perpendicular to the laser polarization. 
As the laser's intensity increases, the scaling of the anomalous current with $n$ 
deviates drastically from the linear response theory, i.e., $\mathbf{J}_\mathbf{\Omega}  \propto  n$. 
Owing to the large linear response from the Berry curvature, 
the anomalous current dominates at  lower intensity. 
However, the ratio of the anomalous current from higher to lower topological charge decreases, which leads
the saturation of the anomalous current relatively at lower intensity. 
On the other hand, normal current -- current parallel to the laser polarization -- scales with the topological charge, and therefore  remains separated in magnitude at relatively higher intensity. 
As the intensity increases, the optical responses from  multi-Weyl semimetals transit from a 
linear regime to 
a nonlinear regime, and  from a perturbative regime to a nonperturbative regime.  
High-harmonic spectroscopy is used 
to probe the distinct and interesting features of the currents in  multi-Weyl semimetals. 
It has been observed that the harmonic yield and the energy cutoff  of the higher-order 
harmonics increase drastically as $n$ increases. 
Moreover,  the polarization of the emitted harmonics encodes 
information about the phase and magnitude of the Berry curvature's components. 
It can be anticipated that the topological charge of the multi-Weyl semimetal 
can be characterized by the polarization of the emitted harmonics in an all-optical way.

\section*{ACKNOWLEDGMENTS}
G. D. acknowledges support from the Science and Engineering Research Board (SERB) India 
(Project No. MTR/2021/000138).  

\newpage
\bibliography{solid_HHG}

\begin{thebibliography}{73}%
\makeatletter
\providecommand \@ifxundefined [1]{%
 \@ifx{#1\undefined}
}%
\providecommand \@ifnum [1]{%
 \ifnum #1\expandafter \@firstoftwo
 \else \expandafter \@secondoftwo
 \fi
}%
\providecommand \@ifx [1]{%
 \ifx #1\expandafter \@firstoftwo
 \else \expandafter \@secondoftwo
 \fi
}%
\providecommand \natexlab [1]{#1}%
\providecommand \enquote  [1]{``#1''}%
\providecommand \bibnamefont  [1]{#1}%
\providecommand \bibfnamefont [1]{#1}%
\providecommand \citenamefont [1]{#1}%
\providecommand \href@noop [0]{\@secondoftwo}%
\providecommand \href [0]{\begingroup \@sanitize@url \@href}%
\providecommand \@href[1]{\@@startlink{#1}\@@href}%
\providecommand \@@href[1]{\endgroup#1\@@endlink}%
\providecommand \@sanitize@url [0]{\catcode `\\12\catcode `\$12\catcode
  `\&12\catcode `\#12\catcode `\^12\catcode `\_12\catcode `\%12\relax}%
\providecommand \@@startlink[1]{}%
\providecommand \@@endlink[0]{}%
\providecommand \url  [0]{\begingroup\@sanitize@url \@url }%
\providecommand \@url [1]{\endgroup\@href {#1}{\urlprefix }}%
\providecommand \urlprefix  [0]{URL }%
\providecommand \Eprint [0]{\href }%
\providecommand \doibase [0]{http://dx.doi.org/}%
\providecommand \selectlanguage [0]{\@gobble}%
\providecommand \bibinfo  [0]{\@secondoftwo}%
\providecommand \bibfield  [0]{\@secondoftwo}%
\providecommand \translation [1]{[#1]}%
\providecommand \BibitemOpen [0]{}%
\providecommand \bibitemStop [0]{}%
\providecommand \bibitemNoStop [0]{.\EOS\space}%
\providecommand \EOS [0]{\spacefactor3000\relax}%
\providecommand \BibitemShut  [1]{\csname bibitem#1\endcsname}%
\let\auto@bib@innerbib\@empty
\bibitem [{\citenamefont {Armitage}\ \emph {et~al.}(2018)\citenamefont
  {Armitage}, \citenamefont {Mele},\ and\ \citenamefont
  {Vishwanath}}]{armitage2018weyl}%
  \BibitemOpen
  \bibfield  {author} {\bibinfo {author} {\bibfnamefont {N.}~\bibnamefont
  {Armitage}}, \bibinfo {author} {\bibfnamefont {E.}~\bibnamefont {Mele}}, \
  and\ \bibinfo {author} {\bibfnamefont {A.}~\bibnamefont {Vishwanath}},\
  }\href@noop {} {\bibfield  {journal} {\bibinfo  {journal} {Revs. Mod. Phys.}\
  }\textbf {\bibinfo {volume} {90}},\ \bibinfo {pages} {015001} (\bibinfo
  {year} {2018})}\BibitemShut {NoStop}%
\bibitem [{\citenamefont {Yan}\ and\ \citenamefont
  {Felser}(2017)}]{yan2017topological}%
  \BibitemOpen
  \bibfield  {author} {\bibinfo {author} {\bibfnamefont {B.}~\bibnamefont
  {Yan}}\ and\ \bibinfo {author} {\bibfnamefont {C.}~\bibnamefont {Felser}},\
  }\href@noop {} {\bibfield  {journal} {\bibinfo  {journal} {Annual Review of
  Condensed Matter Physics}\ }\textbf {\bibinfo {volume} {8}},\ \bibinfo
  {pages} {337} (\bibinfo {year} {2017})}\BibitemShut {NoStop}%
\bibitem [{\citenamefont {Vazifeh}\ and\ \citenamefont
  {Franz}(2013)}]{vazifeh2013electromagnetic}%
  \BibitemOpen
  \bibfield  {author} {\bibinfo {author} {\bibfnamefont {M.}~\bibnamefont
  {Vazifeh}}\ and\ \bibinfo {author} {\bibfnamefont {M.}~\bibnamefont
  {Franz}},\ }\href@noop {} {\bibfield  {journal} {\bibinfo  {journal}
  {Physical Review Letters}\ }\textbf {\bibinfo {volume} {111}},\ \bibinfo
  {pages} {027201} (\bibinfo {year} {2013})}\BibitemShut {NoStop}%
\bibitem [{\citenamefont {Li}\ \emph {et~al.}(2016)\citenamefont {Li},
  \citenamefont {Kharzeev}, \citenamefont {Zhang}, \citenamefont {Huang},
  \citenamefont {Pletikosi{\'c}}, \citenamefont {Fedorov}, \citenamefont
  {Zhong}, \citenamefont {Schneeloch}, \citenamefont {Gu},\ and\ \citenamefont
  {Valla}}]{li2016chiral}%
  \BibitemOpen
  \bibfield  {author} {\bibinfo {author} {\bibfnamefont {Q.}~\bibnamefont
  {Li}}, \bibinfo {author} {\bibfnamefont {D.~E.}\ \bibnamefont {Kharzeev}},
  \bibinfo {author} {\bibfnamefont {C.}~\bibnamefont {Zhang}}, \bibinfo
  {author} {\bibfnamefont {Y.}~\bibnamefont {Huang}}, \bibinfo {author}
  {\bibfnamefont {I.}~\bibnamefont {Pletikosi{\'c}}}, \bibinfo {author}
  {\bibfnamefont {A.}~\bibnamefont {Fedorov}}, \bibinfo {author} {\bibfnamefont
  {R.}~\bibnamefont {Zhong}}, \bibinfo {author} {\bibfnamefont
  {J.}~\bibnamefont {Schneeloch}}, \bibinfo {author} {\bibfnamefont
  {G.}~\bibnamefont {Gu}}, \ and\ \bibinfo {author} {\bibfnamefont
  {T.}~\bibnamefont {Valla}},\ }\href@noop {} {\bibfield  {journal} {\bibinfo
  {journal} {Nature Physics}\ }\textbf {\bibinfo {volume} {12}},\ \bibinfo
  {pages} {550} (\bibinfo {year} {2016})}\BibitemShut {NoStop}%
\bibitem [{\citenamefont {Kaushik}\ \emph {et~al.}(2019)\citenamefont
  {Kaushik}, \citenamefont {Kharzeev},\ and\ \citenamefont
  {Philip}}]{kaushik2019chiral}%
  \BibitemOpen
  \bibfield  {author} {\bibinfo {author} {\bibfnamefont {S.}~\bibnamefont
  {Kaushik}}, \bibinfo {author} {\bibfnamefont {D.~E.}\ \bibnamefont
  {Kharzeev}}, \ and\ \bibinfo {author} {\bibfnamefont {E.~J.}\ \bibnamefont
  {Philip}},\ }\href@noop {} {\bibfield  {journal} {\bibinfo  {journal}
  {Physical Review B}\ }\textbf {\bibinfo {volume} {99}},\ \bibinfo {pages}
  {075150} (\bibinfo {year} {2019})}\BibitemShut {NoStop}%
\bibitem [{\citenamefont {de~Juan}\ \emph {et~al.}(2017)\citenamefont
  {de~Juan}, \citenamefont {Grushin}, \citenamefont {Morimoto},\ and\
  \citenamefont {Moore}}]{de2017quantized}%
  \BibitemOpen
  \bibfield  {author} {\bibinfo {author} {\bibfnamefont {F.}~\bibnamefont
  {de~Juan}}, \bibinfo {author} {\bibfnamefont {A.~G.}\ \bibnamefont
  {Grushin}}, \bibinfo {author} {\bibfnamefont {T.}~\bibnamefont {Morimoto}}, \
  and\ \bibinfo {author} {\bibfnamefont {J.~E.}\ \bibnamefont {Moore}},\
  }\href@noop {} {\bibfield  {journal} {\bibinfo  {journal} {Nature
  communications}\ }\textbf {\bibinfo {volume} {8}},\ \bibinfo {pages} {1}
  (\bibinfo {year} {2017})}\BibitemShut {NoStop}%
\bibitem [{\citenamefont {Ma}\ \emph {et~al.}(2017)\citenamefont {Ma},
  \citenamefont {Xu}, \citenamefont {Chan}, \citenamefont {Zhang},
  \citenamefont {Chang}, \citenamefont {Lin}, \citenamefont {Xie},
  \citenamefont {Palacios}, \citenamefont {Lin}, \citenamefont {Jia} \emph
  {et~al.}}]{ma2017direct}%
  \BibitemOpen
  \bibfield  {author} {\bibinfo {author} {\bibfnamefont {Q.}~\bibnamefont
  {Ma}}, \bibinfo {author} {\bibfnamefont {S.-Y.}\ \bibnamefont {Xu}}, \bibinfo
  {author} {\bibfnamefont {C.-K.}\ \bibnamefont {Chan}}, \bibinfo {author}
  {\bibfnamefont {C.-L.}\ \bibnamefont {Zhang}}, \bibinfo {author}
  {\bibfnamefont {G.}~\bibnamefont {Chang}}, \bibinfo {author} {\bibfnamefont
  {Y.}~\bibnamefont {Lin}}, \bibinfo {author} {\bibfnamefont {W.}~\bibnamefont
  {Xie}}, \bibinfo {author} {\bibfnamefont {T.}~\bibnamefont {Palacios}},
  \bibinfo {author} {\bibfnamefont {H.}~\bibnamefont {Lin}}, \bibinfo {author}
  {\bibfnamefont {S.}~\bibnamefont {Jia}},  \emph {et~al.},\ }\href@noop {}
  {\bibfield  {journal} {\bibinfo  {journal} {Nature Physics}\ }\textbf
  {\bibinfo {volume} {13}},\ \bibinfo {pages} {842} (\bibinfo {year}
  {2017})}\BibitemShut {NoStop}%
\bibitem [{\citenamefont {Rees}\ \emph {et~al.}(2020)\citenamefont {Rees},
  \citenamefont {Manna}, \citenamefont {Lu}, \citenamefont {Morimoto},
  \citenamefont {Borrmann}, \citenamefont {Felser}, \citenamefont {Moore},
  \citenamefont {Torchinsky},\ and\ \citenamefont
  {Orenstein}}]{rees2020helicity}%
  \BibitemOpen
  \bibfield  {author} {\bibinfo {author} {\bibfnamefont {D.}~\bibnamefont
  {Rees}}, \bibinfo {author} {\bibfnamefont {K.}~\bibnamefont {Manna}},
  \bibinfo {author} {\bibfnamefont {B.}~\bibnamefont {Lu}}, \bibinfo {author}
  {\bibfnamefont {T.}~\bibnamefont {Morimoto}}, \bibinfo {author}
  {\bibfnamefont {H.}~\bibnamefont {Borrmann}}, \bibinfo {author}
  {\bibfnamefont {C.}~\bibnamefont {Felser}}, \bibinfo {author} {\bibfnamefont
  {J.}~\bibnamefont {Moore}}, \bibinfo {author} {\bibfnamefont {D.~H.}\
  \bibnamefont {Torchinsky}}, \ and\ \bibinfo {author} {\bibfnamefont
  {J.}~\bibnamefont {Orenstein}},\ }\href@noop {} {\bibfield  {journal}
  {\bibinfo  {journal} {Science advances}\ }\textbf {\bibinfo {volume} {6}},\
  \bibinfo {pages} {eaba0509} (\bibinfo {year} {2020})}\BibitemShut {NoStop}%
\bibitem [{\citenamefont {Le}\ and\ \citenamefont
  {Sun}(2021)}]{le2021topology}%
  \BibitemOpen
  \bibfield  {author} {\bibinfo {author} {\bibfnamefont {C.}~\bibnamefont
  {Le}}\ and\ \bibinfo {author} {\bibfnamefont {Y.}~\bibnamefont {Sun}},\
  }\href@noop {} {\bibfield  {journal} {\bibinfo  {journal} {Journal of
  Physics: Condensed Matter}\ }\textbf {\bibinfo {volume} {33}},\ \bibinfo
  {pages} {503003} (\bibinfo {year} {2021})}\BibitemShut {NoStop}%
\bibitem [{\citenamefont {Shekhar}\ \emph {et~al.}(2018)\citenamefont
  {Shekhar}, \citenamefont {Kumar}, \citenamefont {Grinenko}, \citenamefont
  {Singh}, \citenamefont {Sarkar}, \citenamefont {Luetkens}, \citenamefont
  {Wu}, \citenamefont {Zhang}, \citenamefont {Komarek}, \citenamefont {Kampert}
  \emph {et~al.}}]{shekhar2018anomalous}%
  \BibitemOpen
  \bibfield  {author} {\bibinfo {author} {\bibfnamefont {C.}~\bibnamefont
  {Shekhar}}, \bibinfo {author} {\bibfnamefont {N.}~\bibnamefont {Kumar}},
  \bibinfo {author} {\bibfnamefont {V.}~\bibnamefont {Grinenko}}, \bibinfo
  {author} {\bibfnamefont {S.}~\bibnamefont {Singh}}, \bibinfo {author}
  {\bibfnamefont {R.}~\bibnamefont {Sarkar}}, \bibinfo {author} {\bibfnamefont
  {H.}~\bibnamefont {Luetkens}}, \bibinfo {author} {\bibfnamefont {S.-C.}\
  \bibnamefont {Wu}}, \bibinfo {author} {\bibfnamefont {Y.}~\bibnamefont
  {Zhang}}, \bibinfo {author} {\bibfnamefont {A.~C.}\ \bibnamefont {Komarek}},
  \bibinfo {author} {\bibfnamefont {E.}~\bibnamefont {Kampert}},  \emph
  {et~al.},\ }\href@noop {} {\bibfield  {journal} {\bibinfo  {journal}
  {Proceedings of the National Academy of Sciences}\ }\textbf {\bibinfo
  {volume} {115}},\ \bibinfo {pages} {9140} (\bibinfo {year}
  {2018})}\BibitemShut {NoStop}%
\bibitem [{\citenamefont {Meng}\ \emph {et~al.}(2019)\citenamefont {Meng},
  \citenamefont {Wu}, \citenamefont {Qiu}, \citenamefont {Wang}, \citenamefont
  {Liu}, \citenamefont {Xia}, \citenamefont {Yuan}, \citenamefont {Chang},\
  and\ \citenamefont {Tian}}]{meng2019large}%
  \BibitemOpen
  \bibfield  {author} {\bibinfo {author} {\bibfnamefont {B.}~\bibnamefont
  {Meng}}, \bibinfo {author} {\bibfnamefont {H.}~\bibnamefont {Wu}}, \bibinfo
  {author} {\bibfnamefont {Y.}~\bibnamefont {Qiu}}, \bibinfo {author}
  {\bibfnamefont {C.}~\bibnamefont {Wang}}, \bibinfo {author} {\bibfnamefont
  {Y.}~\bibnamefont {Liu}}, \bibinfo {author} {\bibfnamefont {Z.}~\bibnamefont
  {Xia}}, \bibinfo {author} {\bibfnamefont {S.}~\bibnamefont {Yuan}}, \bibinfo
  {author} {\bibfnamefont {H.}~\bibnamefont {Chang}}, \ and\ \bibinfo {author}
  {\bibfnamefont {Z.}~\bibnamefont {Tian}},\ }\href@noop {} {\bibfield
  {journal} {\bibinfo  {journal} {APL Materials}\ }\textbf {\bibinfo {volume}
  {7}},\ \bibinfo {pages} {051110} (\bibinfo {year} {2019})}\BibitemShut
  {NoStop}%
\bibitem [{\citenamefont {Yang}\ \emph {et~al.}(2015)\citenamefont {Yang},
  \citenamefont {Pan},\ and\ \citenamefont {Zhang}}]{yang2015chirality}%
  \BibitemOpen
  \bibfield  {author} {\bibinfo {author} {\bibfnamefont {S.~A.}\ \bibnamefont
  {Yang}}, \bibinfo {author} {\bibfnamefont {H.}~\bibnamefont {Pan}}, \ and\
  \bibinfo {author} {\bibfnamefont {F.}~\bibnamefont {Zhang}},\ }\href@noop {}
  {\bibfield  {journal} {\bibinfo  {journal} {Physical Review Letters}\
  }\textbf {\bibinfo {volume} {115}},\ \bibinfo {pages} {156603} (\bibinfo
  {year} {2015})}\BibitemShut {NoStop}%
\bibitem [{\citenamefont {Burkov}(2014)}]{burkov2014chiral}%
  \BibitemOpen
  \bibfield  {author} {\bibinfo {author} {\bibfnamefont {A.}~\bibnamefont
  {Burkov}},\ }\href@noop {} {\bibfield  {journal} {\bibinfo  {journal}
  {Physical Review Letters}\ }\textbf {\bibinfo {volume} {113}},\ \bibinfo
  {pages} {247203} (\bibinfo {year} {2014})}\BibitemShut {NoStop}%
\bibitem [{\citenamefont {Trescher}\ \emph {et~al.}(2015)\citenamefont
  {Trescher}, \citenamefont {Sbierski}, \citenamefont {Brouwer},\ and\
  \citenamefont {Bergholtz}}]{trescher2015quantum}%
  \BibitemOpen
  \bibfield  {author} {\bibinfo {author} {\bibfnamefont {M.}~\bibnamefont
  {Trescher}}, \bibinfo {author} {\bibfnamefont {B.}~\bibnamefont {Sbierski}},
  \bibinfo {author} {\bibfnamefont {P.~W.}\ \bibnamefont {Brouwer}}, \ and\
  \bibinfo {author} {\bibfnamefont {E.~J.}\ \bibnamefont {Bergholtz}},\
  }\href@noop {} {\bibfield  {journal} {\bibinfo  {journal} {Physical Review
  B}\ }\textbf {\bibinfo {volume} {91}},\ \bibinfo {pages} {115135} (\bibinfo
  {year} {2015})}\BibitemShut {NoStop}%
\bibitem [{\citenamefont {Kim}\ \emph {et~al.}(2013)\citenamefont {Kim},
  \citenamefont {Kim}, \citenamefont {Wang}, \citenamefont {Sasaki},
  \citenamefont {Satoh}, \citenamefont {Ohnishi}, \citenamefont {Kitaura},
  \citenamefont {Yang},\ and\ \citenamefont {Li}}]{kim2013dirac}%
  \BibitemOpen
  \bibfield  {author} {\bibinfo {author} {\bibfnamefont {H.-J.}\ \bibnamefont
  {Kim}}, \bibinfo {author} {\bibfnamefont {K.-S.}\ \bibnamefont {Kim}},
  \bibinfo {author} {\bibfnamefont {J.-F.}\ \bibnamefont {Wang}}, \bibinfo
  {author} {\bibfnamefont {M.}~\bibnamefont {Sasaki}}, \bibinfo {author}
  {\bibfnamefont {N.}~\bibnamefont {Satoh}}, \bibinfo {author} {\bibfnamefont
  {A.}~\bibnamefont {Ohnishi}}, \bibinfo {author} {\bibfnamefont
  {M.}~\bibnamefont {Kitaura}}, \bibinfo {author} {\bibfnamefont
  {M.}~\bibnamefont {Yang}}, \ and\ \bibinfo {author} {\bibfnamefont
  {L.}~\bibnamefont {Li}},\ }\href@noop {} {\bibfield  {journal} {\bibinfo
  {journal} {Physical Review Letters}\ }\textbf {\bibinfo {volume} {111}},\
  \bibinfo {pages} {246603} (\bibinfo {year} {2013})}\BibitemShut {NoStop}%
\bibitem [{\citenamefont {Lv}\ \emph {et~al.}(2021{\natexlab{a}})\citenamefont
  {Lv}, \citenamefont {Qian},\ and\ \citenamefont {Ding}}]{lv2021experimental}%
  \BibitemOpen
  \bibfield  {author} {\bibinfo {author} {\bibfnamefont {B.}~\bibnamefont
  {Lv}}, \bibinfo {author} {\bibfnamefont {T.}~\bibnamefont {Qian}}, \ and\
  \bibinfo {author} {\bibfnamefont {H.}~\bibnamefont {Ding}},\ }\href@noop {}
  {\bibfield  {journal} {\bibinfo  {journal} {Reviews of Modern Physics}\
  }\textbf {\bibinfo {volume} {93}},\ \bibinfo {pages} {025002} (\bibinfo
  {year} {2021}{\natexlab{a}})}\BibitemShut {NoStop}%
\bibitem [{\citenamefont {Lv}\ \emph {et~al.}(2021{\natexlab{b}})\citenamefont
  {Lv}, \citenamefont {Xu}, \citenamefont {Han}, \citenamefont {Zhang},
  \citenamefont {Han}, \citenamefont {Zhou}, \citenamefont {Yao}, \citenamefont
  {Liu}, \citenamefont {Lu}, \citenamefont {Weng} \emph {et~al.}}]{lv2021high}%
  \BibitemOpen
  \bibfield  {author} {\bibinfo {author} {\bibfnamefont {Y.-Y.}\ \bibnamefont
  {Lv}}, \bibinfo {author} {\bibfnamefont {J.}~\bibnamefont {Xu}}, \bibinfo
  {author} {\bibfnamefont {S.}~\bibnamefont {Han}}, \bibinfo {author}
  {\bibfnamefont {C.}~\bibnamefont {Zhang}}, \bibinfo {author} {\bibfnamefont
  {Y.}~\bibnamefont {Han}}, \bibinfo {author} {\bibfnamefont {J.}~\bibnamefont
  {Zhou}}, \bibinfo {author} {\bibfnamefont {S.-H.}\ \bibnamefont {Yao}},
  \bibinfo {author} {\bibfnamefont {X.-P.}\ \bibnamefont {Liu}}, \bibinfo
  {author} {\bibfnamefont {M.-H.}\ \bibnamefont {Lu}}, \bibinfo {author}
  {\bibfnamefont {H.}~\bibnamefont {Weng}},  \emph {et~al.},\ }\href@noop {}
  {\bibfield  {journal} {\bibinfo  {journal} {Nature Communications}\ }\textbf
  {\bibinfo {volume} {12}},\ \bibinfo {pages} {1} (\bibinfo {year}
  {2021}{\natexlab{b}})}\BibitemShut {NoStop}%
\bibitem [{\citenamefont {Bao}\ \emph {et~al.}(2021)\citenamefont {Bao},
  \citenamefont {Tang}, \citenamefont {Sun},\ and\ \citenamefont
  {Zhou}}]{bao2021light}%
  \BibitemOpen
  \bibfield  {author} {\bibinfo {author} {\bibfnamefont {C.}~\bibnamefont
  {Bao}}, \bibinfo {author} {\bibfnamefont {P.}~\bibnamefont {Tang}}, \bibinfo
  {author} {\bibfnamefont {D.}~\bibnamefont {Sun}}, \ and\ \bibinfo {author}
  {\bibfnamefont {S.}~\bibnamefont {Zhou}},\ }\href@noop {} {\bibfield
  {journal} {\bibinfo  {journal} {Nature Reviews Physics}\ ,\ \bibinfo {pages}
  {1}} (\bibinfo {year} {2021})}\BibitemShut {NoStop}%
\bibitem [{\citenamefont {Orenstein}\ \emph {et~al.}(2021)\citenamefont
  {Orenstein}, \citenamefont {Moore}, \citenamefont {Morimoto}, \citenamefont
  {Torchinsky}, \citenamefont {Harter},\ and\ \citenamefont
  {Hsieh}}]{orenstein2021topology}%
  \BibitemOpen
  \bibfield  {author} {\bibinfo {author} {\bibfnamefont {J.}~\bibnamefont
  {Orenstein}}, \bibinfo {author} {\bibfnamefont {J.}~\bibnamefont {Moore}},
  \bibinfo {author} {\bibfnamefont {T.}~\bibnamefont {Morimoto}}, \bibinfo
  {author} {\bibfnamefont {D.}~\bibnamefont {Torchinsky}}, \bibinfo {author}
  {\bibfnamefont {J.}~\bibnamefont {Harter}}, \ and\ \bibinfo {author}
  {\bibfnamefont {D.}~\bibnamefont {Hsieh}},\ }\href@noop {} {\bibfield
  {journal} {\bibinfo  {journal} {Annual Review of Condensed Matter Physics}\
  }\textbf {\bibinfo {volume} {12}},\ \bibinfo {pages} {247} (\bibinfo {year}
  {2021})}\BibitemShut {NoStop}%
\bibitem [{\citenamefont {Dantas}\ \emph {et~al.}(2021)\citenamefont {Dantas},
  \citenamefont {Wang}, \citenamefont {Sur{\'o}wka},\ and\ \citenamefont
  {Oka}}]{dantas2021nonperturbative}%
  \BibitemOpen
  \bibfield  {author} {\bibinfo {author} {\bibfnamefont {R.~M.}\ \bibnamefont
  {Dantas}}, \bibinfo {author} {\bibfnamefont {Z.}~\bibnamefont {Wang}},
  \bibinfo {author} {\bibfnamefont {P.}~\bibnamefont {Sur{\'o}wka}}, \ and\
  \bibinfo {author} {\bibfnamefont {T.}~\bibnamefont {Oka}},\ }\href@noop {}
  {\bibfield  {journal} {\bibinfo  {journal} {Physical Review B}\ }\textbf
  {\bibinfo {volume} {103}},\ \bibinfo {pages} {L201105} (\bibinfo {year}
  {2021})}\BibitemShut {NoStop}%
\bibitem [{\citenamefont {Tamashevich}\ \emph {et~al.}(2022)\citenamefont
  {Tamashevich}, \citenamefont {Villari},\ and\ \citenamefont
  {Ornigotti}}]{tamashevich2022nonlinear}%
  \BibitemOpen
  \bibfield  {author} {\bibinfo {author} {\bibfnamefont {Y.}~\bibnamefont
  {Tamashevich}}, \bibinfo {author} {\bibfnamefont {L.~D.~M.}\ \bibnamefont
  {Villari}}, \ and\ \bibinfo {author} {\bibfnamefont {M.}~\bibnamefont
  {Ornigotti}},\ }\href@noop {} {\bibfield  {journal} {\bibinfo  {journal}
  {Physical Review B}\ }\textbf {\bibinfo {volume} {105}},\ \bibinfo {pages}
  {195102} (\bibinfo {year} {2022})}\BibitemShut {NoStop}%
\bibitem [{\citenamefont {Nathan}\ \emph {et~al.}(2022)\citenamefont {Nathan},
  \citenamefont {Martin},\ and\ \citenamefont
  {Refael}}]{nathan2022topological}%
  \BibitemOpen
  \bibfield  {author} {\bibinfo {author} {\bibfnamefont {F.}~\bibnamefont
  {Nathan}}, \bibinfo {author} {\bibfnamefont {I.}~\bibnamefont {Martin}}, \
  and\ \bibinfo {author} {\bibfnamefont {G.}~\bibnamefont {Refael}},\
  }\href@noop {} {\bibfield  {journal} {\bibinfo  {journal} {Physical Review
  Research}\ }\textbf {\bibinfo {volume} {4}},\ \bibinfo {pages} {043060}
  (\bibinfo {year} {2022})}\BibitemShut {NoStop}%
\bibitem [{\citenamefont {Matsyshyn}\ \emph {et~al.}(2021)\citenamefont
  {Matsyshyn}, \citenamefont {Piazza}, \citenamefont {Moessner},\ and\
  \citenamefont {Sodemann}}]{matsyshyn2021rabi}%
  \BibitemOpen
  \bibfield  {author} {\bibinfo {author} {\bibfnamefont {O.}~\bibnamefont
  {Matsyshyn}}, \bibinfo {author} {\bibfnamefont {F.}~\bibnamefont {Piazza}},
  \bibinfo {author} {\bibfnamefont {R.}~\bibnamefont {Moessner}}, \ and\
  \bibinfo {author} {\bibfnamefont {I.}~\bibnamefont {Sodemann}},\ }\href@noop
  {} {\bibfield  {journal} {\bibinfo  {journal} {Physical Review Letters}\
  }\textbf {\bibinfo {volume} {127}},\ \bibinfo {pages} {126604} (\bibinfo
  {year} {2021})}\BibitemShut {NoStop}%
\bibitem [{\citenamefont {Ahn}\ \emph {et~al.}(2017)\citenamefont {Ahn},
  \citenamefont {Mele},\ and\ \citenamefont {Min}}]{ahn2017optical}%
  \BibitemOpen
  \bibfield  {author} {\bibinfo {author} {\bibfnamefont {S.}~\bibnamefont
  {Ahn}}, \bibinfo {author} {\bibfnamefont {E.}~\bibnamefont {Mele}}, \ and\
  \bibinfo {author} {\bibfnamefont {H.}~\bibnamefont {Min}},\ }\href@noop {}
  {\bibfield  {journal} {\bibinfo  {journal} {Physical Review B}\ }\textbf
  {\bibinfo {volume} {95}},\ \bibinfo {pages} {161112} (\bibinfo {year}
  {2017})}\BibitemShut {NoStop}%
\bibitem [{\citenamefont {Sirica}\ and\ \citenamefont
  {Prasankumar}(2021)}]{sirica2021shaking}%
  \BibitemOpen
  \bibfield  {author} {\bibinfo {author} {\bibfnamefont {N.}~\bibnamefont
  {Sirica}}\ and\ \bibinfo {author} {\bibfnamefont {R.}~\bibnamefont
  {Prasankumar}},\ }\href@noop {} {\bibfield  {journal} {\bibinfo  {journal}
  {Nature Materials}\ }\textbf {\bibinfo {volume} {20}},\ \bibinfo {pages}
  {283} (\bibinfo {year} {2021})}\BibitemShut {NoStop}%
\bibitem [{\citenamefont {Osterhoudt}\ \emph {et~al.}(2019)\citenamefont
  {Osterhoudt}, \citenamefont {Diebel}, \citenamefont {Gray}, \citenamefont
  {Yang}, \citenamefont {Stanco}, \citenamefont {Huang}, \citenamefont {Shen},
  \citenamefont {Ni}, \citenamefont {Moll}, \citenamefont {Ran} \emph
  {et~al.}}]{osterhoudt2019colossal}%
  \BibitemOpen
  \bibfield  {author} {\bibinfo {author} {\bibfnamefont {G.~B.}\ \bibnamefont
  {Osterhoudt}}, \bibinfo {author} {\bibfnamefont {L.~K.}\ \bibnamefont
  {Diebel}}, \bibinfo {author} {\bibfnamefont {M.~J.}\ \bibnamefont {Gray}},
  \bibinfo {author} {\bibfnamefont {X.}~\bibnamefont {Yang}}, \bibinfo {author}
  {\bibfnamefont {J.}~\bibnamefont {Stanco}}, \bibinfo {author} {\bibfnamefont
  {X.}~\bibnamefont {Huang}}, \bibinfo {author} {\bibfnamefont
  {B.}~\bibnamefont {Shen}}, \bibinfo {author} {\bibfnamefont {N.}~\bibnamefont
  {Ni}}, \bibinfo {author} {\bibfnamefont {P.~J.}\ \bibnamefont {Moll}},
  \bibinfo {author} {\bibfnamefont {Y.}~\bibnamefont {Ran}},  \emph {et~al.},\
  }\href@noop {} {\bibfield  {journal} {\bibinfo  {journal} {Nature materials}\
  }\textbf {\bibinfo {volume} {18}},\ \bibinfo {pages} {471} (\bibinfo {year}
  {2019})}\BibitemShut {NoStop}%
\bibitem [{\citenamefont {Gao}\ \emph {et~al.}(2020)\citenamefont {Gao},
  \citenamefont {Kaushik}, \citenamefont {Philip}, \citenamefont {Li},
  \citenamefont {Qin}, \citenamefont {Liu}, \citenamefont {Zhang},
  \citenamefont {Su}, \citenamefont {Chen}, \citenamefont {Weng} \emph
  {et~al.}}]{gao2020chiral}%
  \BibitemOpen
  \bibfield  {author} {\bibinfo {author} {\bibfnamefont {Y.}~\bibnamefont
  {Gao}}, \bibinfo {author} {\bibfnamefont {S.}~\bibnamefont {Kaushik}},
  \bibinfo {author} {\bibfnamefont {E.}~\bibnamefont {Philip}}, \bibinfo
  {author} {\bibfnamefont {Z.}~\bibnamefont {Li}}, \bibinfo {author}
  {\bibfnamefont {Y.}~\bibnamefont {Qin}}, \bibinfo {author} {\bibfnamefont
  {Y.}~\bibnamefont {Liu}}, \bibinfo {author} {\bibfnamefont {W.}~\bibnamefont
  {Zhang}}, \bibinfo {author} {\bibfnamefont {Y.}~\bibnamefont {Su}}, \bibinfo
  {author} {\bibfnamefont {X.}~\bibnamefont {Chen}}, \bibinfo {author}
  {\bibfnamefont {H.}~\bibnamefont {Weng}},  \emph {et~al.},\ }\href@noop {}
  {\bibfield  {journal} {\bibinfo  {journal} {Nature communications}\ }\textbf
  {\bibinfo {volume} {11}},\ \bibinfo {pages} {1} (\bibinfo {year}
  {2020})}\BibitemShut {NoStop}%
\bibitem [{\citenamefont {Lv}\ \emph {et~al.}(2015)\citenamefont {Lv},
  \citenamefont {Weng}, \citenamefont {Fu}, \citenamefont {Wang}, \citenamefont
  {Miao}, \citenamefont {Ma}, \citenamefont {Richard}, \citenamefont {Huang},
  \citenamefont {Zhao}, \citenamefont {Chen} \emph
  {et~al.}}]{lv2015experimental}%
  \BibitemOpen
  \bibfield  {author} {\bibinfo {author} {\bibfnamefont {B.}~\bibnamefont
  {Lv}}, \bibinfo {author} {\bibfnamefont {H.}~\bibnamefont {Weng}}, \bibinfo
  {author} {\bibfnamefont {B.}~\bibnamefont {Fu}}, \bibinfo {author}
  {\bibfnamefont {X.~P.}\ \bibnamefont {Wang}}, \bibinfo {author}
  {\bibfnamefont {H.}~\bibnamefont {Miao}}, \bibinfo {author} {\bibfnamefont
  {J.}~\bibnamefont {Ma}}, \bibinfo {author} {\bibfnamefont {P.}~\bibnamefont
  {Richard}}, \bibinfo {author} {\bibfnamefont {X.}~\bibnamefont {Huang}},
  \bibinfo {author} {\bibfnamefont {L.}~\bibnamefont {Zhao}}, \bibinfo {author}
  {\bibfnamefont {G.}~\bibnamefont {Chen}},  \emph {et~al.},\ }\href@noop {}
  {\bibfield  {journal} {\bibinfo  {journal} {Physical Review X}\ }\textbf
  {\bibinfo {volume} {5}},\ \bibinfo {pages} {031013} (\bibinfo {year}
  {2015})}\BibitemShut {NoStop}%
\bibitem [{\citenamefont {Xu}\ \emph {et~al.}(2015)\citenamefont {Xu},
  \citenamefont {Alidoust}, \citenamefont {Belopolski}, \citenamefont {Yuan},
  \citenamefont {Bian}, \citenamefont {Chang}, \citenamefont {Zheng},
  \citenamefont {Strocov}, \citenamefont {Sanchez}, \citenamefont {Chang} \emph
  {et~al.}}]{xu2015discovery}%
  \BibitemOpen
  \bibfield  {author} {\bibinfo {author} {\bibfnamefont {S.-Y.}\ \bibnamefont
  {Xu}}, \bibinfo {author} {\bibfnamefont {N.}~\bibnamefont {Alidoust}},
  \bibinfo {author} {\bibfnamefont {I.}~\bibnamefont {Belopolski}}, \bibinfo
  {author} {\bibfnamefont {Z.}~\bibnamefont {Yuan}}, \bibinfo {author}
  {\bibfnamefont {G.}~\bibnamefont {Bian}}, \bibinfo {author} {\bibfnamefont
  {T.-R.}\ \bibnamefont {Chang}}, \bibinfo {author} {\bibfnamefont
  {H.}~\bibnamefont {Zheng}}, \bibinfo {author} {\bibfnamefont {V.~N.}\
  \bibnamefont {Strocov}}, \bibinfo {author} {\bibfnamefont {D.~S.}\
  \bibnamefont {Sanchez}}, \bibinfo {author} {\bibfnamefont {G.}~\bibnamefont
  {Chang}},  \emph {et~al.},\ }\href@noop {} {\bibfield  {journal} {\bibinfo
  {journal} {Nature Physics}\ }\textbf {\bibinfo {volume} {11}},\ \bibinfo
  {pages} {748} (\bibinfo {year} {2015})}\BibitemShut {NoStop}%
\bibitem [{\citenamefont {Morali}\ \emph {et~al.}(2019)\citenamefont {Morali},
  \citenamefont {Batabyal}, \citenamefont {Nag}, \citenamefont {Liu},
  \citenamefont {Xu}, \citenamefont {Sun}, \citenamefont {Yan}, \citenamefont
  {Felser}, \citenamefont {Avraham},\ and\ \citenamefont
  {Beidenkopf}}]{morali2019fermi}%
  \BibitemOpen
  \bibfield  {author} {\bibinfo {author} {\bibfnamefont {N.}~\bibnamefont
  {Morali}}, \bibinfo {author} {\bibfnamefont {R.}~\bibnamefont {Batabyal}},
  \bibinfo {author} {\bibfnamefont {P.~K.}\ \bibnamefont {Nag}}, \bibinfo
  {author} {\bibfnamefont {E.}~\bibnamefont {Liu}}, \bibinfo {author}
  {\bibfnamefont {Q.}~\bibnamefont {Xu}}, \bibinfo {author} {\bibfnamefont
  {Y.}~\bibnamefont {Sun}}, \bibinfo {author} {\bibfnamefont {B.}~\bibnamefont
  {Yan}}, \bibinfo {author} {\bibfnamefont {C.}~\bibnamefont {Felser}},
  \bibinfo {author} {\bibfnamefont {N.}~\bibnamefont {Avraham}}, \ and\
  \bibinfo {author} {\bibfnamefont {H.}~\bibnamefont {Beidenkopf}},\
  }\href@noop {} {\bibfield  {journal} {\bibinfo  {journal} {Science}\ }\textbf
  {\bibinfo {volume} {365}},\ \bibinfo {pages} {1286} (\bibinfo {year}
  {2019})}\BibitemShut {NoStop}%
\bibitem [{\citenamefont {Liu}\ \emph {et~al.}(2019)\citenamefont {Liu},
  \citenamefont {Liang}, \citenamefont {Liu}, \citenamefont {Xu}, \citenamefont
  {Li}, \citenamefont {Chen}, \citenamefont {Pei}, \citenamefont {Shi},
  \citenamefont {Mo}, \citenamefont {Dudin} \emph {et~al.}}]{liu2019magnetic}%
  \BibitemOpen
  \bibfield  {author} {\bibinfo {author} {\bibfnamefont {D.}~\bibnamefont
  {Liu}}, \bibinfo {author} {\bibfnamefont {A.}~\bibnamefont {Liang}}, \bibinfo
  {author} {\bibfnamefont {E.}~\bibnamefont {Liu}}, \bibinfo {author}
  {\bibfnamefont {Q.}~\bibnamefont {Xu}}, \bibinfo {author} {\bibfnamefont
  {Y.}~\bibnamefont {Li}}, \bibinfo {author} {\bibfnamefont {C.}~\bibnamefont
  {Chen}}, \bibinfo {author} {\bibfnamefont {D.}~\bibnamefont {Pei}}, \bibinfo
  {author} {\bibfnamefont {W.}~\bibnamefont {Shi}}, \bibinfo {author}
  {\bibfnamefont {S.}~\bibnamefont {Mo}}, \bibinfo {author} {\bibfnamefont
  {P.}~\bibnamefont {Dudin}},  \emph {et~al.},\ }\href@noop {} {\bibfield
  {journal} {\bibinfo  {journal} {Science}\ }\textbf {\bibinfo {volume}
  {365}},\ \bibinfo {pages} {1282} (\bibinfo {year} {2019})}\BibitemShut
  {NoStop}%
\bibitem [{\citenamefont {Belopolski}\ \emph {et~al.}(2019)\citenamefont
  {Belopolski}, \citenamefont {Manna}, \citenamefont {Sanchez}, \citenamefont
  {Chang}, \citenamefont {Ernst}, \citenamefont {Yin}, \citenamefont {Zhang},
  \citenamefont {Cochran}, \citenamefont {Shumiya}, \citenamefont {Zheng} \emph
  {et~al.}}]{belopolski2019discovery}%
  \BibitemOpen
  \bibfield  {author} {\bibinfo {author} {\bibfnamefont {I.}~\bibnamefont
  {Belopolski}}, \bibinfo {author} {\bibfnamefont {K.}~\bibnamefont {Manna}},
  \bibinfo {author} {\bibfnamefont {D.~S.}\ \bibnamefont {Sanchez}}, \bibinfo
  {author} {\bibfnamefont {G.}~\bibnamefont {Chang}}, \bibinfo {author}
  {\bibfnamefont {B.}~\bibnamefont {Ernst}}, \bibinfo {author} {\bibfnamefont
  {J.}~\bibnamefont {Yin}}, \bibinfo {author} {\bibfnamefont {S.~S.}\
  \bibnamefont {Zhang}}, \bibinfo {author} {\bibfnamefont {T.}~\bibnamefont
  {Cochran}}, \bibinfo {author} {\bibfnamefont {N.}~\bibnamefont {Shumiya}},
  \bibinfo {author} {\bibfnamefont {H.}~\bibnamefont {Zheng}},  \emph
  {et~al.},\ }\href@noop {} {\bibfield  {journal} {\bibinfo  {journal}
  {Science}\ }\textbf {\bibinfo {volume} {365}},\ \bibinfo {pages} {1278}
  (\bibinfo {year} {2019})}\BibitemShut {NoStop}%
\bibitem [{\citenamefont {Fang}\ \emph {et~al.}(2012)\citenamefont {Fang},
  \citenamefont {Gilbert}, \citenamefont {Dai},\ and\ \citenamefont
  {Bernevig}}]{fang2012multi}%
  \BibitemOpen
  \bibfield  {author} {\bibinfo {author} {\bibfnamefont {C.}~\bibnamefont
  {Fang}}, \bibinfo {author} {\bibfnamefont {M.~J.}\ \bibnamefont {Gilbert}},
  \bibinfo {author} {\bibfnamefont {X.}~\bibnamefont {Dai}}, \ and\ \bibinfo
  {author} {\bibfnamefont {B.~A.}\ \bibnamefont {Bernevig}},\ }\href@noop {}
  {\bibfield  {journal} {\bibinfo  {journal} {Physical Review Letters}\
  }\textbf {\bibinfo {volume} {108}},\ \bibinfo {pages} {266802} (\bibinfo
  {year} {2012})}\BibitemShut {NoStop}%
\bibitem [{\citenamefont {Huang}\ \emph {et~al.}(2016)\citenamefont {Huang},
  \citenamefont {Xu}, \citenamefont {Belopolski}, \citenamefont {Lee},
  \citenamefont {Chang}, \citenamefont {Chang}, \citenamefont {Wang},
  \citenamefont {Alidoust}, \citenamefont {Bian}, \citenamefont {Neupane} \emph
  {et~al.}}]{huang2016new}%
  \BibitemOpen
  \bibfield  {author} {\bibinfo {author} {\bibfnamefont {S.-M.}\ \bibnamefont
  {Huang}}, \bibinfo {author} {\bibfnamefont {S.-Y.}\ \bibnamefont {Xu}},
  \bibinfo {author} {\bibfnamefont {I.}~\bibnamefont {Belopolski}}, \bibinfo
  {author} {\bibfnamefont {C.-C.}\ \bibnamefont {Lee}}, \bibinfo {author}
  {\bibfnamefont {G.}~\bibnamefont {Chang}}, \bibinfo {author} {\bibfnamefont
  {T.-R.}\ \bibnamefont {Chang}}, \bibinfo {author} {\bibfnamefont
  {B.}~\bibnamefont {Wang}}, \bibinfo {author} {\bibfnamefont {N.}~\bibnamefont
  {Alidoust}}, \bibinfo {author} {\bibfnamefont {G.}~\bibnamefont {Bian}},
  \bibinfo {author} {\bibfnamefont {M.}~\bibnamefont {Neupane}},  \emph
  {et~al.},\ }\href@noop {} {\bibfield  {journal} {\bibinfo  {journal}
  {Proceedings of the National Academy of Sciences}\ }\textbf {\bibinfo
  {volume} {113}},\ \bibinfo {pages} {1180} (\bibinfo {year}
  {2016})}\BibitemShut {NoStop}%
\bibitem [{\citenamefont {Xu}\ \emph {et~al.}(2011)\citenamefont {Xu},
  \citenamefont {Weng}, \citenamefont {Wang}, \citenamefont {Dai},\ and\
  \citenamefont {Fang}}]{xu2011chern}%
  \BibitemOpen
  \bibfield  {author} {\bibinfo {author} {\bibfnamefont {G.}~\bibnamefont
  {Xu}}, \bibinfo {author} {\bibfnamefont {H.}~\bibnamefont {Weng}}, \bibinfo
  {author} {\bibfnamefont {Z.}~\bibnamefont {Wang}}, \bibinfo {author}
  {\bibfnamefont {X.}~\bibnamefont {Dai}}, \ and\ \bibinfo {author}
  {\bibfnamefont {Z.}~\bibnamefont {Fang}},\ }\href@noop {} {\bibfield
  {journal} {\bibinfo  {journal} {Physical Review Letters}\ }\textbf {\bibinfo
  {volume} {107}},\ \bibinfo {pages} {186806} (\bibinfo {year}
  {2011})}\BibitemShut {NoStop}%
\bibitem [{\citenamefont {Liu}\ and\ \citenamefont
  {Zunger}(2017)}]{liu2017predicted}%
  \BibitemOpen
  \bibfield  {author} {\bibinfo {author} {\bibfnamefont {Q.}~\bibnamefont
  {Liu}}\ and\ \bibinfo {author} {\bibfnamefont {A.}~\bibnamefont {Zunger}},\
  }\href@noop {} {\bibfield  {journal} {\bibinfo  {journal} {Physical Review
  X}\ }\textbf {\bibinfo {volume} {7}},\ \bibinfo {pages} {021019} (\bibinfo
  {year} {2017})}\BibitemShut {NoStop}%
\bibitem [{\citenamefont {Roy}\ and\ \citenamefont
  {Narayan}(2022)}]{roy2022non}%
  \BibitemOpen
  \bibfield  {author} {\bibinfo {author} {\bibfnamefont {S.}~\bibnamefont
  {Roy}}\ and\ \bibinfo {author} {\bibfnamefont {A.}~\bibnamefont {Narayan}},\
  }\href@noop {} {\bibfield  {journal} {\bibinfo  {journal} {Journal of
  Physics: Condensed Matter}\ }\textbf {\bibinfo {volume} {34}},\ \bibinfo
  {pages} {385301} (\bibinfo {year} {2022})}\BibitemShut {NoStop}%
\bibitem [{\citenamefont {Yang}\ and\ \citenamefont
  {Nagaosa}(2014)}]{yang2014classification}%
  \BibitemOpen
  \bibfield  {author} {\bibinfo {author} {\bibfnamefont {B.-J.}\ \bibnamefont
  {Yang}}\ and\ \bibinfo {author} {\bibfnamefont {N.}~\bibnamefont {Nagaosa}},\
  }\href@noop {} {\bibfield  {journal} {\bibinfo  {journal} {Nature
  Communications}\ }\textbf {\bibinfo {volume} {5}},\ \bibinfo {pages} {4898}
  (\bibinfo {year} {2014})}\BibitemShut {NoStop}%
\bibitem [{\citenamefont {Nag}\ \emph {et~al.}(2020)\citenamefont {Nag},
  \citenamefont {Menon},\ and\ \citenamefont {Basu}}]{nag2020thermoelectric}%
  \BibitemOpen
  \bibfield  {author} {\bibinfo {author} {\bibfnamefont {T.}~\bibnamefont
  {Nag}}, \bibinfo {author} {\bibfnamefont {A.}~\bibnamefont {Menon}}, \ and\
  \bibinfo {author} {\bibfnamefont {B.}~\bibnamefont {Basu}},\ }\href@noop {}
  {\bibfield  {journal} {\bibinfo  {journal} {Physical Review B}\ }\textbf
  {\bibinfo {volume} {102}},\ \bibinfo {pages} {014307} (\bibinfo {year}
  {2020})}\BibitemShut {NoStop}%
\bibitem [{\citenamefont {Huang}\ \emph {et~al.}(2017)\citenamefont {Huang},
  \citenamefont {Zhou},\ and\ \citenamefont {Shen}}]{huang2017topological}%
  \BibitemOpen
  \bibfield  {author} {\bibinfo {author} {\bibfnamefont {Z.-M.}\ \bibnamefont
  {Huang}}, \bibinfo {author} {\bibfnamefont {J.}~\bibnamefont {Zhou}}, \ and\
  \bibinfo {author} {\bibfnamefont {S.-Q.}\ \bibnamefont {Shen}},\ }\href@noop
  {} {\bibfield  {journal} {\bibinfo  {journal} {Physical Review B}\ }\textbf
  {\bibinfo {volume} {96}},\ \bibinfo {pages} {085201} (\bibinfo {year}
  {2017})}\BibitemShut {NoStop}%
\bibitem [{\citenamefont {Nandy}\ \emph {et~al.}(2021)\citenamefont {Nandy},
  \citenamefont {Zeng},\ and\ \citenamefont {Tewari}}]{nandy2021chiral}%
  \BibitemOpen
  \bibfield  {author} {\bibinfo {author} {\bibfnamefont {S.}~\bibnamefont
  {Nandy}}, \bibinfo {author} {\bibfnamefont {C.}~\bibnamefont {Zeng}}, \ and\
  \bibinfo {author} {\bibfnamefont {S.}~\bibnamefont {Tewari}},\ }\href@noop {}
  {\bibfield  {journal} {\bibinfo  {journal} {Physical Review B}\ }\textbf
  {\bibinfo {volume} {104}},\ \bibinfo {pages} {205124} (\bibinfo {year}
  {2021})}\BibitemShut {NoStop}%
\bibitem [{\citenamefont {Menon}\ and\ \citenamefont
  {Basu}(2020)}]{menon2020anomalous}%
  \BibitemOpen
  \bibfield  {author} {\bibinfo {author} {\bibfnamefont {A.}~\bibnamefont
  {Menon}}\ and\ \bibinfo {author} {\bibfnamefont {B.}~\bibnamefont {Basu}},\
  }\href@noop {} {\bibfield  {journal} {\bibinfo  {journal} {Journal of
  Physics: Condensed Matter}\ }\textbf {\bibinfo {volume} {33}},\ \bibinfo
  {pages} {045602} (\bibinfo {year} {2020})}\BibitemShut {NoStop}%
\bibitem [{\citenamefont {Menon}\ \emph {et~al.}(2021)\citenamefont {Menon},
  \citenamefont {Chattopadhay},\ and\ \citenamefont {Basu}}]{menon2021chiral}%
  \BibitemOpen
  \bibfield  {author} {\bibinfo {author} {\bibfnamefont {A.}~\bibnamefont
  {Menon}}, \bibinfo {author} {\bibfnamefont {S.}~\bibnamefont {Chattopadhay}},
  \ and\ \bibinfo {author} {\bibfnamefont {B.}~\bibnamefont {Basu}},\
  }\href@noop {} {\bibfield  {journal} {\bibinfo  {journal} {Physical Review
  B}\ }\textbf {\bibinfo {volume} {104}},\ \bibinfo {pages} {075129} (\bibinfo
  {year} {2021})}\BibitemShut {NoStop}%
\bibitem [{\citenamefont {Nag}\ and\ \citenamefont
  {Kennes}(2022)}]{nag2022distinct}%
  \BibitemOpen
  \bibfield  {author} {\bibinfo {author} {\bibfnamefont {T.}~\bibnamefont
  {Nag}}\ and\ \bibinfo {author} {\bibfnamefont {D.~M.}\ \bibnamefont
  {Kennes}},\ }\href@noop {} {\bibfield  {journal} {\bibinfo  {journal}
  {Physical Review B}\ }\textbf {\bibinfo {volume} {105}},\ \bibinfo {pages}
  {214307} (\bibinfo {year} {2022})}\BibitemShut {NoStop}%
\bibitem [{\citenamefont {Chen}\ and\ \citenamefont
  {Fiete}(2016)}]{chen2016thermoelectric}%
  \BibitemOpen
  \bibfield  {author} {\bibinfo {author} {\bibfnamefont {Q.}~\bibnamefont
  {Chen}}\ and\ \bibinfo {author} {\bibfnamefont {G.~A.}\ \bibnamefont
  {Fiete}},\ }\href@noop {} {\bibfield  {journal} {\bibinfo  {journal}
  {Physical Review B}\ }\textbf {\bibinfo {volume} {93}},\ \bibinfo {pages}
  {155125} (\bibinfo {year} {2016})}\BibitemShut {NoStop}%
\bibitem [{\citenamefont {Gorbar}\ \emph {et~al.}(2017)\citenamefont {Gorbar},
  \citenamefont {Miransky}, \citenamefont {Shovkovy},\ and\ \citenamefont
  {Sukhachov}}]{gorbar2017anomalous}%
  \BibitemOpen
  \bibfield  {author} {\bibinfo {author} {\bibfnamefont {E.}~\bibnamefont
  {Gorbar}}, \bibinfo {author} {\bibfnamefont {V.}~\bibnamefont {Miransky}},
  \bibinfo {author} {\bibfnamefont {I.}~\bibnamefont {Shovkovy}}, \ and\
  \bibinfo {author} {\bibfnamefont {P.}~\bibnamefont {Sukhachov}},\ }\href@noop
  {} {\bibfield  {journal} {\bibinfo  {journal} {Physical Review B}\ }\textbf
  {\bibinfo {volume} {96}},\ \bibinfo {pages} {155138} (\bibinfo {year}
  {2017})}\BibitemShut {NoStop}%
\bibitem [{\citenamefont {Park}\ \emph {et~al.}(2017)\citenamefont {Park},
  \citenamefont {Woo}, \citenamefont {Mele},\ and\ \citenamefont
  {Min}}]{park2017semiclassical}%
  \BibitemOpen
  \bibfield  {author} {\bibinfo {author} {\bibfnamefont {S.}~\bibnamefont
  {Park}}, \bibinfo {author} {\bibfnamefont {S.}~\bibnamefont {Woo}}, \bibinfo
  {author} {\bibfnamefont {E.}~\bibnamefont {Mele}}, \ and\ \bibinfo {author}
  {\bibfnamefont {H.}~\bibnamefont {Min}},\ }\href@noop {} {\bibfield
  {journal} {\bibinfo  {journal} {Physical Review B}\ }\textbf {\bibinfo
  {volume} {95}},\ \bibinfo {pages} {161113} (\bibinfo {year}
  {2017})}\BibitemShut {NoStop}%
\bibitem [{\citenamefont {Nandy}\ \emph {et~al.}(2019)\citenamefont {Nandy},
  \citenamefont {Manna}, \citenamefont {C{\u{a}}lug{\u{a}}ru},\ and\
  \citenamefont {Roy}}]{nandy2019generalized}%
  \BibitemOpen
  \bibfield  {author} {\bibinfo {author} {\bibfnamefont {S.}~\bibnamefont
  {Nandy}}, \bibinfo {author} {\bibfnamefont {S.}~\bibnamefont {Manna}},
  \bibinfo {author} {\bibfnamefont {D.}~\bibnamefont {C{\u{a}}lug{\u{a}}ru}}, \
  and\ \bibinfo {author} {\bibfnamefont {B.}~\bibnamefont {Roy}},\ }\href@noop
  {} {\bibfield  {journal} {\bibinfo  {journal} {Physical Review B}\ }\textbf
  {\bibinfo {volume} {100}},\ \bibinfo {pages} {235201} (\bibinfo {year}
  {2019})}\BibitemShut {NoStop}%
\bibitem [{\citenamefont {Dantas}\ \emph {et~al.}(2020)\citenamefont {Dantas},
  \citenamefont {Pe{\~n}a-Benitez}, \citenamefont {Roy},\ and\ \citenamefont
  {Sur{\'o}wka}}]{dantas2020non}%
  \BibitemOpen
  \bibfield  {author} {\bibinfo {author} {\bibfnamefont {R.~M.}\ \bibnamefont
  {Dantas}}, \bibinfo {author} {\bibfnamefont {F.}~\bibnamefont
  {Pe{\~n}a-Benitez}}, \bibinfo {author} {\bibfnamefont {B.}~\bibnamefont
  {Roy}}, \ and\ \bibinfo {author} {\bibfnamefont {P.}~\bibnamefont
  {Sur{\'o}wka}},\ }\href@noop {} {\bibfield  {journal} {\bibinfo  {journal}
  {Physical Review Research}\ }\textbf {\bibinfo {volume} {2}},\ \bibinfo
  {pages} {013007} (\bibinfo {year} {2020})}\BibitemShut {NoStop}%
\bibitem [{\citenamefont {Mrudul}\ \emph {et~al.}(2021)\citenamefont {Mrudul},
  \citenamefont {Jim{\'e}nez-Gal{\'a}n}, \citenamefont {Ivanov},\ and\
  \citenamefont {Dixit}}]{mrudul2021light}%
  \BibitemOpen
  \bibfield  {author} {\bibinfo {author} {\bibfnamefont {M.}~\bibnamefont
  {Mrudul}}, \bibinfo {author} {\bibfnamefont {{\'A}.}~\bibnamefont
  {Jim{\'e}nez-Gal{\'a}n}}, \bibinfo {author} {\bibfnamefont {M.}~\bibnamefont
  {Ivanov}}, \ and\ \bibinfo {author} {\bibfnamefont {G.}~\bibnamefont
  {Dixit}},\ }\href@noop {} {\bibfield  {journal} {\bibinfo  {journal}
  {Optica}\ }\textbf {\bibinfo {volume} {8}},\ \bibinfo {pages} {422} (\bibinfo
  {year} {2021})}\BibitemShut {NoStop}%
\bibitem [{\citenamefont {Mrudul}\ and\ \citenamefont
  {Dixit}(2021{\natexlab{a}})}]{mrudul2021high}%
  \BibitemOpen
  \bibfield  {author} {\bibinfo {author} {\bibfnamefont {M.}~\bibnamefont
  {Mrudul}}\ and\ \bibinfo {author} {\bibfnamefont {G.}~\bibnamefont {Dixit}},\
  }\href@noop {} {\bibfield  {journal} {\bibinfo  {journal} {Physical Review
  B}\ }\textbf {\bibinfo {volume} {103}},\ \bibinfo {pages} {094308} (\bibinfo
  {year} {2021}{\natexlab{a}})}\BibitemShut {NoStop}%
\bibitem [{\citenamefont {Yue}\ and\ \citenamefont
  {Gaarde}(2022)}]{yue2022introduction}%
  \BibitemOpen
  \bibfield  {author} {\bibinfo {author} {\bibfnamefont {L.}~\bibnamefont
  {Yue}}\ and\ \bibinfo {author} {\bibfnamefont {M.~B.}\ \bibnamefont
  {Gaarde}},\ }\href@noop {} {\bibfield  {journal} {\bibinfo  {journal} {JOSA
  B}\ }\textbf {\bibinfo {volume} {39}},\ \bibinfo {pages} {535} (\bibinfo
  {year} {2022})}\BibitemShut {NoStop}%
\bibitem [{\citenamefont {Nematollahi}\ \emph {et~al.}(2020)\citenamefont
  {Nematollahi}, \citenamefont {Motlagh}, \citenamefont {Wu}, \citenamefont
  {Ghimire}, \citenamefont {Apalkov},\ and\ \citenamefont
  {Stockman}}]{nematollahi2020topological}%
  \BibitemOpen
  \bibfield  {author} {\bibinfo {author} {\bibfnamefont {F.}~\bibnamefont
  {Nematollahi}}, \bibinfo {author} {\bibfnamefont {S.~A.~O.}\ \bibnamefont
  {Motlagh}}, \bibinfo {author} {\bibfnamefont {J.-S.}\ \bibnamefont {Wu}},
  \bibinfo {author} {\bibfnamefont {R.}~\bibnamefont {Ghimire}}, \bibinfo
  {author} {\bibfnamefont {V.}~\bibnamefont {Apalkov}}, \ and\ \bibinfo
  {author} {\bibfnamefont {M.~I.}\ \bibnamefont {Stockman}},\ }\href@noop {}
  {\bibfield  {journal} {\bibinfo  {journal} {Physical Review B}\ }\textbf
  {\bibinfo {volume} {102}},\ \bibinfo {pages} {125413} (\bibinfo {year}
  {2020})}\BibitemShut {NoStop}%
\bibitem [{\citenamefont {Ngo}\ \emph {et~al.}(2021)\citenamefont {Ngo},
  \citenamefont {Duc}, \citenamefont {Song}, \citenamefont {Meier} \emph
  {et~al.}}]{ngo2021microscopic}%
  \BibitemOpen
  \bibfield  {author} {\bibinfo {author} {\bibfnamefont {C.}~\bibnamefont
  {Ngo}}, \bibinfo {author} {\bibfnamefont {H.~T.}\ \bibnamefont {Duc}},
  \bibinfo {author} {\bibfnamefont {X.}~\bibnamefont {Song}}, \bibinfo {author}
  {\bibfnamefont {T.}~\bibnamefont {Meier}},  \emph {et~al.},\ }\href@noop {}
  {\bibfield  {journal} {\bibinfo  {journal} {Physical Review B}\ }\textbf
  {\bibinfo {volume} {103}},\ \bibinfo {pages} {085201} (\bibinfo {year}
  {2021})}\BibitemShut {NoStop}%
\bibitem [{\citenamefont {Gu}\ and\ \citenamefont
  {Kolesik}(2022)}]{gu2022full}%
  \BibitemOpen
  \bibfield  {author} {\bibinfo {author} {\bibfnamefont {J.}~\bibnamefont
  {Gu}}\ and\ \bibinfo {author} {\bibfnamefont {M.}~\bibnamefont {Kolesik}},\
  }\href@noop {} {\bibfield  {journal} {\bibinfo  {journal} {Physical Review
  A}\ }\textbf {\bibinfo {volume} {106}},\ \bibinfo {pages} {063516} (\bibinfo
  {year} {2022})}\BibitemShut {NoStop}%
\bibitem [{\citenamefont {Bharti}\ \emph {et~al.}(2022)\citenamefont {Bharti},
  \citenamefont {Mrudul},\ and\ \citenamefont {Dixit}}]{bharti2022high}%
  \BibitemOpen
  \bibfield  {author} {\bibinfo {author} {\bibfnamefont {A.}~\bibnamefont
  {Bharti}}, \bibinfo {author} {\bibfnamefont {M.}~\bibnamefont {Mrudul}}, \
  and\ \bibinfo {author} {\bibfnamefont {G.}~\bibnamefont {Dixit}},\
  }\href@noop {} {\bibfield  {journal} {\bibinfo  {journal} {Physical Review
  B}\ }\textbf {\bibinfo {volume} {105}},\ \bibinfo {pages} {155140} (\bibinfo
  {year} {2022})}\BibitemShut {NoStop}%
\bibitem [{Not()}]{NoteX}%
  \BibitemOpen
  \href@noop {} {}\bibinfo {note} {See Supplemental Material at
  http://link.aps.org/supplemental/ for Berry curvature's components for
  multi-Weyl semimetals with different topological charges, Comparison of the
  anomalous current with the current originating solely from the Berry
  curvature, and variation in the anomalous current with respect to the
  distance between the Weyl nodes at laser's intensity $10^{8}$
  W/cm$^2$.}\BibitemShut {Stop}%
\bibitem [{\citenamefont {Xiao}\ \emph {et~al.}(2010)\citenamefont {Xiao},
  \citenamefont {Chang},\ and\ \citenamefont {Niu}}]{xiao2010berry}%
  \BibitemOpen
  \bibfield  {author} {\bibinfo {author} {\bibfnamefont {D.}~\bibnamefont
  {Xiao}}, \bibinfo {author} {\bibfnamefont {M.-C.}\ \bibnamefont {Chang}}, \
  and\ \bibinfo {author} {\bibfnamefont {Q.}~\bibnamefont {Niu}},\ }\href@noop
  {} {\bibfield  {journal} {\bibinfo  {journal} {Revs. Mod. Phys.}\ }\textbf
  {\bibinfo {volume} {82}},\ \bibinfo {pages} {1959} (\bibinfo {year}
  {2010})}\BibitemShut {NoStop}%
\bibitem [{\citenamefont {Avetissian}\ \emph {et~al.}(2022)\citenamefont
  {Avetissian}, \citenamefont {Avetisyan}, \citenamefont {Avchyan},\ and\
  \citenamefont {Mkrtchian}}]{avetissian2022high}%
  \BibitemOpen
  \bibfield  {author} {\bibinfo {author} {\bibfnamefont {H.}~\bibnamefont
  {Avetissian}}, \bibinfo {author} {\bibfnamefont {V.}~\bibnamefont
  {Avetisyan}}, \bibinfo {author} {\bibfnamefont {B.}~\bibnamefont {Avchyan}},
  \ and\ \bibinfo {author} {\bibfnamefont {G.}~\bibnamefont {Mkrtchian}},\
  }\href@noop {} {\bibfield  {journal} {\bibinfo  {journal} {Physical Review
  A}\ }\textbf {\bibinfo {volume} {106}},\ \bibinfo {pages} {033107} (\bibinfo
  {year} {2022})}\BibitemShut {NoStop}%
\bibitem [{\citenamefont {Mrudul}\ \emph {et~al.}(2020)\citenamefont {Mrudul},
  \citenamefont {Tancogne-Dejean}, \citenamefont {Rubio},\ and\ \citenamefont
  {Dixit}}]{mrudul2020high}%
  \BibitemOpen
  \bibfield  {author} {\bibinfo {author} {\bibfnamefont {M.~S.}\ \bibnamefont
  {Mrudul}}, \bibinfo {author} {\bibfnamefont {N.}~\bibnamefont
  {Tancogne-Dejean}}, \bibinfo {author} {\bibfnamefont {A.}~\bibnamefont
  {Rubio}}, \ and\ \bibinfo {author} {\bibfnamefont {G.}~\bibnamefont
  {Dixit}},\ }\href@noop {} {\bibfield  {journal} {\bibinfo  {journal} {npj
  Computational Materials}\ }\textbf {\bibinfo {volume} {6}},\ \bibinfo {pages}
  {1} (\bibinfo {year} {2020})}\BibitemShut {NoStop}%
\bibitem [{\citenamefont {Schubert}\ \emph {et~al.}(2014)\citenamefont
  {Schubert}, \citenamefont {Hohenleutner}, \citenamefont {Langer},
  \citenamefont {Urbanek}, \citenamefont {Lange}, \citenamefont {Huttner},
  \citenamefont {Golde}, \citenamefont {Meier}, \citenamefont {Kira},
  \citenamefont {Koch},\ and\ \citenamefont {Huber}}]{schubert2014sub}%
  \BibitemOpen
  \bibfield  {author} {\bibinfo {author} {\bibfnamefont {O.}~\bibnamefont
  {Schubert}}, \bibinfo {author} {\bibfnamefont {M.}~\bibnamefont
  {Hohenleutner}}, \bibinfo {author} {\bibfnamefont {F.}~\bibnamefont
  {Langer}}, \bibinfo {author} {\bibfnamefont {B.}~\bibnamefont {Urbanek}},
  \bibinfo {author} {\bibfnamefont {C.}~\bibnamefont {Lange}}, \bibinfo
  {author} {\bibfnamefont {U.}~\bibnamefont {Huttner}}, \bibinfo {author}
  {\bibfnamefont {D.}~\bibnamefont {Golde}}, \bibinfo {author} {\bibfnamefont
  {T.}~\bibnamefont {Meier}}, \bibinfo {author} {\bibfnamefont
  {M.}~\bibnamefont {Kira}}, \bibinfo {author} {\bibfnamefont {S.~W.}\
  \bibnamefont {Koch}}, \ and\ \bibinfo {author} {\bibfnamefont
  {R.}~\bibnamefont {Huber}},\ }\href@noop {} {\bibfield  {journal} {\bibinfo
  {journal} {Nature Photonics}\ }\textbf {\bibinfo {volume} {8}},\ \bibinfo
  {pages} {119} (\bibinfo {year} {2014})}\BibitemShut {NoStop}%
\bibitem [{\citenamefont {Mrudul}\ and\ \citenamefont
  {Dixit}(2021{\natexlab{b}})}]{mrudul2021controlling}%
  \BibitemOpen
  \bibfield  {author} {\bibinfo {author} {\bibfnamefont {M.}~\bibnamefont
  {Mrudul}}\ and\ \bibinfo {author} {\bibfnamefont {G.}~\bibnamefont {Dixit}},\
  }\href@noop {} {\bibfield  {journal} {\bibinfo  {journal} {Journal of Physics
  B: Atomic, Molecular and Optical Physics}\ }\textbf {\bibinfo {volume}
  {54}},\ \bibinfo {pages} {224001} (\bibinfo {year}
  {2021}{\natexlab{b}})}\BibitemShut {NoStop}%
\bibitem [{\citenamefont {Hohenleutner}\ \emph {et~al.}(2015)\citenamefont
  {Hohenleutner}, \citenamefont {Langer}, \citenamefont {Schubert},
  \citenamefont {Knorr}, \citenamefont {Huttner}, \citenamefont {Koch},
  \citenamefont {Kira},\ and\ \citenamefont {Huber}}]{hohenleutner2015real}%
  \BibitemOpen
  \bibfield  {author} {\bibinfo {author} {\bibfnamefont {M.}~\bibnamefont
  {Hohenleutner}}, \bibinfo {author} {\bibfnamefont {F.}~\bibnamefont
  {Langer}}, \bibinfo {author} {\bibfnamefont {O.}~\bibnamefont {Schubert}},
  \bibinfo {author} {\bibfnamefont {M.}~\bibnamefont {Knorr}}, \bibinfo
  {author} {\bibfnamefont {U.}~\bibnamefont {Huttner}}, \bibinfo {author}
  {\bibfnamefont {S.~W.}\ \bibnamefont {Koch}}, \bibinfo {author}
  {\bibfnamefont {M.}~\bibnamefont {Kira}}, \ and\ \bibinfo {author}
  {\bibfnamefont {R.}~\bibnamefont {Huber}},\ }\href@noop {} {\bibfield
  {journal} {\bibinfo  {journal} {Nature}\ }\textbf {\bibinfo {volume} {523}},\
  \bibinfo {pages} {572} (\bibinfo {year} {2015})}\BibitemShut {NoStop}%
\bibitem [{\citenamefont {Zaks}\ \emph {et~al.}(2012)\citenamefont {Zaks},
  \citenamefont {Liu},\ and\ \citenamefont {Sherwin}}]{zaks2012experimental}%
  \BibitemOpen
  \bibfield  {author} {\bibinfo {author} {\bibfnamefont {B.}~\bibnamefont
  {Zaks}}, \bibinfo {author} {\bibfnamefont {R.~B.}\ \bibnamefont {Liu}}, \
  and\ \bibinfo {author} {\bibfnamefont {M.~S.}\ \bibnamefont {Sherwin}},\
  }\href@noop {} {\bibfield  {journal} {\bibinfo  {journal} {Nature}\ }\textbf
  {\bibinfo {volume} {483}},\ \bibinfo {pages} {580} (\bibinfo {year}
  {2012})}\BibitemShut {NoStop}%
\bibitem [{\citenamefont {Pattanayak}\ \emph {et~al.}(2020)\citenamefont
  {Pattanayak}, \citenamefont {Mrudul},\ and\ \citenamefont
  {Dixit}}]{pattanayak2020influence}%
  \BibitemOpen
  \bibfield  {author} {\bibinfo {author} {\bibfnamefont {A.}~\bibnamefont
  {Pattanayak}}, \bibinfo {author} {\bibfnamefont {M.~S.}\ \bibnamefont
  {Mrudul}}, \ and\ \bibinfo {author} {\bibfnamefont {G.}~\bibnamefont
  {Dixit}},\ }\href@noop {} {\bibfield  {journal} {\bibinfo  {journal}
  {Physical Review A}\ }\textbf {\bibinfo {volume} {101}},\ \bibinfo {pages}
  {013404} (\bibinfo {year} {2020})}\BibitemShut {NoStop}%
\bibitem [{\citenamefont {Mrudul}\ \emph {et~al.}(2019)\citenamefont {Mrudul},
  \citenamefont {Pattanayak}, \citenamefont {Ivanov},\ and\ \citenamefont
  {Dixit}}]{pattanayak2019direct}%
  \BibitemOpen
  \bibfield  {author} {\bibinfo {author} {\bibfnamefont {M.~S.}\ \bibnamefont
  {Mrudul}}, \bibinfo {author} {\bibfnamefont {A.}~\bibnamefont {Pattanayak}},
  \bibinfo {author} {\bibfnamefont {M.}~\bibnamefont {Ivanov}}, \ and\ \bibinfo
  {author} {\bibfnamefont {G.}~\bibnamefont {Dixit}},\ }\href@noop {}
  {\bibfield  {journal} {\bibinfo  {journal} {Physical Review A}\ }\textbf
  {\bibinfo {volume} {100}},\ \bibinfo {pages} {043420} (\bibinfo {year}
  {2019})}\BibitemShut {NoStop}%
\bibitem [{\citenamefont {Pattanayak}\ \emph {et~al.}(2022)\citenamefont
  {Pattanayak}, \citenamefont {Pujari},\ and\ \citenamefont
  {Dixit}}]{pattanayak2022role}%
  \BibitemOpen
  \bibfield  {author} {\bibinfo {author} {\bibfnamefont {A.}~\bibnamefont
  {Pattanayak}}, \bibinfo {author} {\bibfnamefont {S.}~\bibnamefont {Pujari}},
  \ and\ \bibinfo {author} {\bibfnamefont {G.}~\bibnamefont {Dixit}},\
  }\href@noop {} {\bibfield  {journal} {\bibinfo  {journal} {Scientific
  Reports}\ }\textbf {\bibinfo {volume} {12}},\ \bibinfo {pages} {6722}
  (\bibinfo {year} {2022})}\BibitemShut {NoStop}%
\bibitem [{\citenamefont {Rana}\ and\ \citenamefont
  {Dixit}(2022)}]{rana2022probing}%
  \BibitemOpen
  \bibfield  {author} {\bibinfo {author} {\bibfnamefont {N.}~\bibnamefont
  {Rana}}\ and\ \bibinfo {author} {\bibfnamefont {G.}~\bibnamefont {Dixit}},\
  }\href@noop {} {\bibfield  {journal} {\bibinfo  {journal} {Physical Review
  A}\ }\textbf {\bibinfo {volume} {106}},\ \bibinfo {pages} {053116} (\bibinfo
  {year} {2022})}\BibitemShut {NoStop}%
\bibitem [{\citenamefont {Rana}\ \emph {et~al.}(2022)\citenamefont {Rana},
  \citenamefont {Mrudul}, \citenamefont {Kartashov}, \citenamefont {Ivanov},\
  and\ \citenamefont {Dixit}}]{rana2022high}%
  \BibitemOpen
  \bibfield  {author} {\bibinfo {author} {\bibfnamefont {N.}~\bibnamefont
  {Rana}}, \bibinfo {author} {\bibfnamefont {M.}~\bibnamefont {Mrudul}},
  \bibinfo {author} {\bibfnamefont {D.}~\bibnamefont {Kartashov}}, \bibinfo
  {author} {\bibfnamefont {M.}~\bibnamefont {Ivanov}}, \ and\ \bibinfo {author}
  {\bibfnamefont {G.}~\bibnamefont {Dixit}},\ }\href@noop {} {\bibfield
  {journal} {\bibinfo  {journal} {Physical Review B}\ }\textbf {\bibinfo
  {volume} {106}},\ \bibinfo {pages} {064303} (\bibinfo {year}
  {2022})}\BibitemShut {NoStop}%
\bibitem [{\citenamefont {Baykusheva}\ \emph {et~al.}(2021)\citenamefont
  {Baykusheva}, \citenamefont {Chac{\'o}n}, \citenamefont {Lu}, \citenamefont
  {Bailey}, \citenamefont {Sobota}, \citenamefont {Soifer}, \citenamefont
  {Kirchmann}, \citenamefont {Rotundu}, \citenamefont {Uher}, \citenamefont
  {Heinz} \emph {et~al.}}]{baykusheva2021all}%
  \BibitemOpen
  \bibfield  {author} {\bibinfo {author} {\bibfnamefont {D.}~\bibnamefont
  {Baykusheva}}, \bibinfo {author} {\bibfnamefont {A.}~\bibnamefont
  {Chac{\'o}n}}, \bibinfo {author} {\bibfnamefont {J.}~\bibnamefont {Lu}},
  \bibinfo {author} {\bibfnamefont {T.~P.}\ \bibnamefont {Bailey}}, \bibinfo
  {author} {\bibfnamefont {J.~A.}\ \bibnamefont {Sobota}}, \bibinfo {author}
  {\bibfnamefont {H.}~\bibnamefont {Soifer}}, \bibinfo {author} {\bibfnamefont
  {P.~S.}\ \bibnamefont {Kirchmann}}, \bibinfo {author} {\bibfnamefont
  {C.}~\bibnamefont {Rotundu}}, \bibinfo {author} {\bibfnamefont
  {C.}~\bibnamefont {Uher}}, \bibinfo {author} {\bibfnamefont {T.~F.}\
  \bibnamefont {Heinz}},  \emph {et~al.},\ }\href@noop {} {\bibfield  {journal}
  {\bibinfo  {journal} {Nano Letters}\ }\textbf {\bibinfo {volume} {21}},\
  \bibinfo {pages} {8970} (\bibinfo {year} {2021})}\BibitemShut {NoStop}%
\bibitem [{\citenamefont {Kovalev}\ \emph {et~al.}(2020)\citenamefont
  {Kovalev}, \citenamefont {Dantas}, \citenamefont {Germanskiy}, \citenamefont
  {Deinert}, \citenamefont {Green}, \citenamefont {Ilyakov}, \citenamefont
  {Awari}, \citenamefont {Chen}, \citenamefont {Bawatna}, \citenamefont {Ling}
  \emph {et~al.}}]{kovalev2020non}%
  \BibitemOpen
  \bibfield  {author} {\bibinfo {author} {\bibfnamefont {S.}~\bibnamefont
  {Kovalev}}, \bibinfo {author} {\bibfnamefont {R.~M.}\ \bibnamefont {Dantas}},
  \bibinfo {author} {\bibfnamefont {S.}~\bibnamefont {Germanskiy}}, \bibinfo
  {author} {\bibfnamefont {J.-C.}\ \bibnamefont {Deinert}}, \bibinfo {author}
  {\bibfnamefont {B.}~\bibnamefont {Green}}, \bibinfo {author} {\bibfnamefont
  {I.}~\bibnamefont {Ilyakov}}, \bibinfo {author} {\bibfnamefont
  {N.}~\bibnamefont {Awari}}, \bibinfo {author} {\bibfnamefont
  {M.}~\bibnamefont {Chen}}, \bibinfo {author} {\bibfnamefont {M.}~\bibnamefont
  {Bawatna}}, \bibinfo {author} {\bibfnamefont {J.}~\bibnamefont {Ling}},
  \emph {et~al.},\ }\href@noop {} {\bibfield  {journal} {\bibinfo  {journal}
  {Nature communications}\ }\textbf {\bibinfo {volume} {11}},\ \bibinfo {pages}
  {2451} (\bibinfo {year} {2020})}\BibitemShut {NoStop}%
\bibitem [{\citenamefont {Cheng}\ \emph {et~al.}(2020)\citenamefont {Cheng},
  \citenamefont {Kanda}, \citenamefont {Ikeda}, \citenamefont {Matsuda},
  \citenamefont {Xia}, \citenamefont {Schumann}, \citenamefont {Stemmer},
  \citenamefont {Itatani}, \citenamefont {Armitage},\ and\ \citenamefont
  {Matsunaga}}]{cheng2020efficient}%
  \BibitemOpen
  \bibfield  {author} {\bibinfo {author} {\bibfnamefont {B.}~\bibnamefont
  {Cheng}}, \bibinfo {author} {\bibfnamefont {N.}~\bibnamefont {Kanda}},
  \bibinfo {author} {\bibfnamefont {T.~N.}\ \bibnamefont {Ikeda}}, \bibinfo
  {author} {\bibfnamefont {T.}~\bibnamefont {Matsuda}}, \bibinfo {author}
  {\bibfnamefont {P.}~\bibnamefont {Xia}}, \bibinfo {author} {\bibfnamefont
  {T.}~\bibnamefont {Schumann}}, \bibinfo {author} {\bibfnamefont
  {S.}~\bibnamefont {Stemmer}}, \bibinfo {author} {\bibfnamefont
  {J.}~\bibnamefont {Itatani}}, \bibinfo {author} {\bibfnamefont
  {N.}~\bibnamefont {Armitage}}, \ and\ \bibinfo {author} {\bibfnamefont
  {R.}~\bibnamefont {Matsunaga}},\ }\href@noop {} {\bibfield  {journal}
  {\bibinfo  {journal} {Physical Review Letters}\ }\textbf {\bibinfo {volume}
  {124}},\ \bibinfo {pages} {117402} (\bibinfo {year} {2020})}\BibitemShut
  {NoStop}%
\bibitem [{\citenamefont {Bai}\ \emph {et~al.}(2021)\citenamefont {Bai},
  \citenamefont {Fei}, \citenamefont {Wang}, \citenamefont {Li}, \citenamefont
  {Li}, \citenamefont {Song}, \citenamefont {Li}, \citenamefont {Xu},\ and\
  \citenamefont {Liu}}]{bai2021high}%
  \BibitemOpen
  \bibfield  {author} {\bibinfo {author} {\bibfnamefont {Y.}~\bibnamefont
  {Bai}}, \bibinfo {author} {\bibfnamefont {F.}~\bibnamefont {Fei}}, \bibinfo
  {author} {\bibfnamefont {S.}~\bibnamefont {Wang}}, \bibinfo {author}
  {\bibfnamefont {N.}~\bibnamefont {Li}}, \bibinfo {author} {\bibfnamefont
  {X.}~\bibnamefont {Li}}, \bibinfo {author} {\bibfnamefont {F.}~\bibnamefont
  {Song}}, \bibinfo {author} {\bibfnamefont {R.}~\bibnamefont {Li}}, \bibinfo
  {author} {\bibfnamefont {Z.}~\bibnamefont {Xu}}, \ and\ \bibinfo {author}
  {\bibfnamefont {P.}~\bibnamefont {Liu}},\ }\href@noop {} {\bibfield
  {journal} {\bibinfo  {journal} {Nature Physics}\ }\textbf {\bibinfo {volume}
  {17}},\ \bibinfo {pages} {311} (\bibinfo {year} {2021})}\BibitemShut
  {NoStop}%
\end{thebibliography}%

\end{document}